%% file: paper_eprint.tex
\documentclass[twocolumn,showpacs,aps,floatfix,prd]{revtex4}

\usepackage{graphicx}
\usepackage{dcolumn}
\usepackage{amsmath}
\usepackage{epsfig}
\usepackage{hhline}
\usepackage{soul}
\usepackage{bm}
\usepackage{tabularx,pifont,multirow}
\usepackage{colordvi}
\usepackage{longtable}
\usepackage{psfrag}
\RequirePackage{xspace}

\newcommand{\BABARPubYear}     {05}
\newcommand{\BABARPubNumber}   {022}
\newcommand{\SLACPubNumber} {11343}

\input pubboard/babarsym

\newcommand{\bi}{\begin{itemize}}
\newcommand{\ei}{\end{itemize}}
\newcommand{\ben}{\begin{enumerate}}
\newcommand{\een}{\end{enumerate}}
\newcommand{\bc}{\begin{center}}
\newcommand{\ec}{\end{center}}
\newcommand{\bt}{\begin{table}}
\newcommand{\et}{\end{table}}
\newcommand{\be}{\begin{equation}}
\newcommand{\eeq}{\end{equation}}
\newcommand{\ba}{\begin{eqnarray}}
\newcommand{\ea}{\end{eqnarray}}
\newcommand{\hs}{\hspace}
\newcommand{\vs}{\vspace}
\newcommand{\la}{\ifmmode {\leftarrow} \else {$\leftarrow$}\fi}
\newcommand{\Ra}{\ifmmode {\Rightarrow} \else {$\Rightarrow$}\fi}
\newcommand{\La}{\ifmmode {\Leftarrow} \else {$\Leftarrow$}\fi}
\newcommand{\Lra}{\ifmmode {\Longrightarrow} \else {$\Longrightarrow$}\fi}
\newcommand{\Lla}{\ifmmode {\Longleftarrow} \else {$\Longleftarrow$\fi}}
\newcommand{\Llra}{\ifmmode {\Longleftrightarrow} \else {$\Longleftrightarrow$\fi}}
\newcommand{\Lk}{\ifmmode {{\cal L}} \else {${\cal L}$}\fi}
\newcommand{\Wt}{\ifmmode {{\cal W}} \else {${\cal W}$}\fi}
\newcommand{\Br}{\ifmmode {{\cal B}} \else {${\cal B}$}\fi}
\newcommand{\N}{\ifmmode {{\cal N}} \else {${\cal N}$}\fi}
\newcommand{\G}{\ifmmode {{\cal G}} \else {${\cal G}$}\fi}
\newcommand{\E}{\ifmmode {{\cal E}} \else {${\cal E}$}\fi}

\newcommand{\tBz}{\ifmmode {\tau_{\Bz}} \else { $\tau_{\Bz}$ } \fi }
\newcommand{\tBu}{\ifmmode {\tau_{\Bub}} \else { $\tau_{\Bub}$ } \fi }
\newcommand{\BtoDs}{\mbox{$\Bbar^0\rightarrow D^{*+} \ell^- \bar{\nu}_\ell$}}
\newcommand{\BtoDss}{\mbox{$B\rightarrow D^{*} \pi \ell^- \bar{\nu_\ell}$}}

\newcommand{\pstar}{\ifmmode {\pi^{+}_{soft}} \else {$\pi^{+}_{soft}$} \fi }
\newcommand{\psoft}{\ifmmode {\pi^{+}_{s}} \else {$\pi^{+}_{s}$} \fi }
\newcommand{\plab}{\ifmmode{p} \else {$p$} \fi}
\newcommand{\ks}{\ifmmode{k^*} \else {$k^*$} \fi}
\newcommand{\mnusq}{\ifmmode{{{\cal M}^2_\nu}} \else {${{\cal M}_{\nu}}^2$}\fi} 
\newcommand{\DTau}{\ifmmode {\Delta \tau} \else {$\Delta \tau$}\fi}
\newcommand{\dmd}{\ifmmode {\Delta m_d} \else {$\Delta m_d$}\fi}
\newcommand{\deltams}{\ifmmode {\Delta m_s} \else {$\Delta m_s$}\fi}
\newcommand{\ggcc}{\ifmmode {GeV^2/c^4} \else {$GeV^2/c^4$}\fi}
\def\BpBm {\ensuremath{B^+ {\kern -0.16em \Bub}}}

\def\ptl{\ensuremath{{\cal P}\ell}}
\def\ctl{\ensuremath{{\cal C}\ell}}
\def\cdl{\ensuremath{{\cal D}\ell}}

\def\tDe{\ensuremath{\tau_{D_e}}}
\def\dz{\ensuremath{\Delta z}}

\def\st{\ensuremath{\sigma_{\Delta t}}}

\def\chid{\ensuremath{\chi_d}}

\long\def\inst#1{\par\nobreak\kern 4pt\nobreak
  {\it #1}\par\vskip 10pt plus 3pt minus 3pt}

\begin{document}

\begin{flushleft}
SLAC-PUB-\SLACPubNumber \\
\babar-PUB-\BABARPubYear/\BABARPubNumber \\
\end{flushleft}

\begin{center}
\title{\large \bf
\boldmath
\Large Measurement of the {\boldmath \Bzb} Lifetime and the
{\boldmath \BzBzb} Oscillation Frequency Using Partially Reconstructed 
{\boldmath $\Bzb\rightarrow D^{*+} \ell^- {\overline\nu_\ell}~$} Decays

\vskip 5mm
The \babar\ Collaboration
\mbox{ }\\
\end{center}
} 
\input pubboard/authors_may2005.tex

\date{\today}

\begin{abstract}
We present a simultaneous measurement of the \Bzb\ lifetime $\tBz$ and 
$\BzBzb$ oscillation frequency \deltamd. We use a sample of about 
50\,000 partially reconstructed \BtoDs\ decays identified with the 
\babar\ detector at the \pep2\ \epem\ storage ring at SLAC. The flavor 
of the other $B$ meson in the event is determined from the charge of 
another high-momentum lepton. The results are
\ba
\nonumber \tBz &=& (1.504 \pm 0.013 ~\mathrm{(stat)} ~^{+0.018}_{-0.013} ~\mathrm{(syst))~ps}, \\
\nonumber \dmd &=& (0.511 \pm 0.007 ~\mathrm{(stat)} ~^{+0.007}_{-0.006} ~\mathrm{(syst))~ps}^{-1}.
\ea
\end{abstract}


\maketitle



\setcounter{footnote}{0}


\section{INTRODUCTION}
\label{sec:Introduction}
The time evolution of \Bzb\ mesons 
is governed by the overall decay
rate $\Gamma(\Bzb) = 1/\tBz$ and by the mass difference \deltamd\ of the two mass
eigenstates.
A precise determination of $\Gamma(\Bzb)$ reduces the systematic error
on the parameter $\Vcb$
of the Cabibbo-Kobayashi-Maskawa quark mixing matrix~\cite{ref:CKM}.
The parameter $|V_{td}V_{tb}^*|$ enters the box diagram that is responsible for \BzBzb\ oscillations and
can be determined from a measurement of \deltamd, although with sizable theoretical uncertainties. 
\par
We present a measurement of $\tBz$ and \deltamd\ using
\BtoDs\ decays 
\cite{ref:footnote1}
selected from a sample of about 88 million \BB\ events recorded by the \babar\ 
detector at the \pep2\ asymmetric-energy \epem\ storage ring, operated at or near the
\FourS\ resonance. \BB\ pairs from the \FourS\ decay move along the beam axis with a nominal 
Lorentz boost $\langle \beta\gamma \rangle = 0.56$, so that the vertices from the two $B$ decay points are
separated on average by about 260 $\mu$m.
The \BzBzb\ system is produced in a coherent $P$-wave state, so that flavor oscillation is
measurable only relative to the decay of the first $B$ meson. Mixed (unmixed) events are selected 
by the observation of two equal (opposite) flavor $B$ meson decays.
The probabilities of observing
mixed (${\cal S}^{-}$) or unmixed (${\cal S}^{+}$) events as a function of the proper time 
difference \deltat\ between decays are
\ba
\label{eq:pdf} {\cal S^{\pm}}(\deltat) = \frac{e^{-|\deltat|/\tBz}}{4\tBz} (1 \pm {\cal D} \cos(\deltamd \deltat) ),
\ea
where the dilution factor ${\cal D}$ is related to the fraction $w$ of events with wrong 
flavor assignment by the relation  ${\cal D}=1-2w$ and \deltat\ is computed from the distance between
the two vertices projected along the beam direction.
\par

\section{THE \babar\ DETECTOR AND DATASET}
\label{sec:babar}
We have analyzed a data sample of 81 \invfb\ collected by \babar\ on the \FourS\ resonance, a sample of
9.6 \invfb\ collected 40~MeV below the resonance to study the continuum background, 
and a sample of simulated \BB\ events corresponding to about three times 
the size of the data sample. 
The simulated events are processed through the same analysis 
chain as the real data.
\babar\ is a multi-purpose detector, described in detail in Ref.~\cite{ref:babar}. The momentum of
charged particles is measured by 
the tracking system, which consists of a silicon vertex tracker (SVT) 
and a drift chamber (DCH) in a 1.5-T magnetic field.
The positions of points along the trajectories of charged tracks measured with the SVT are used for vertex reconstruction and for measuring the momentum of charged particles, including those particles with low transverse momentum that do not reach the DCH due to bending in the magnetic field.
The energy loss in the SVT is used to discriminate low-momentum
pions from electrons. 
Higher-energy electrons are identified from the ratio of the energy of their associated shower in 
the electromagnetic calorimeter (EMC) to their momentum, the transverse profile of the 
shower, the energy loss in the DCH, and the information from the Cherenkov detector 
(DIRC). The electron identification efficiency is about $90\%$, and 
the hadron misidentification probability is less than $1\%$. 
Muons are identified on the basis of the energy deposited
in the EMC and the penetration in the instrumented flux return (IFR) of the superconducting coil, 
which contains resistive plate chambers interspersed with iron. 
Muon candidates compatible with the kaon hypothesis in the DIRC are rejected. The muon identification 
efficiency is about $60\%$, and the hadron misidentification rate is about 2\%.
\par

\section{ANALYSIS METHOD}
\label{sec:Analysis}
\subsection{Selection of \boldmath \BtoDs\ decays \label{s:sele}}

We select events that have more than four charged tracks.
We reduce the contamination from light-quark production in continuum events by
requiring the normalized Fox-Wolfram second moment~\cite{ref:FW} to be less than 0.5.
We select \BtoDs\ events with partial reconstruction of the decay $\dsp \ra \psoft \Dz$,
using only the charged lepton from the \Bzb\ decay and the soft pion (\psoft) from the \dsp decay.
The \Dz\ decay is not reconstructed, resulting in high selection efficiency.
\babar\ has already published two measurements of $\tBz$ \cite{ref:t1,ref:t2} and a measurement of 
$\sin(2\beta + \gamma)$~\cite{ref:sin2bg} based on partial 
reconstruction of $B$ decays. This technique
was originally applied to \BtoDs\ decays by ARGUS \cite{ref:ARGUS}, and then used
by CLEO \cite{ref:CLEO}, DELPHI \cite{ref:DELPHI}, and OPAL \cite{ref:OPAL}. 
\par
\begin{figure}[t]
\vs{-.4cm}
\begin{center}
\includegraphics[width=9cm,height=12cm]{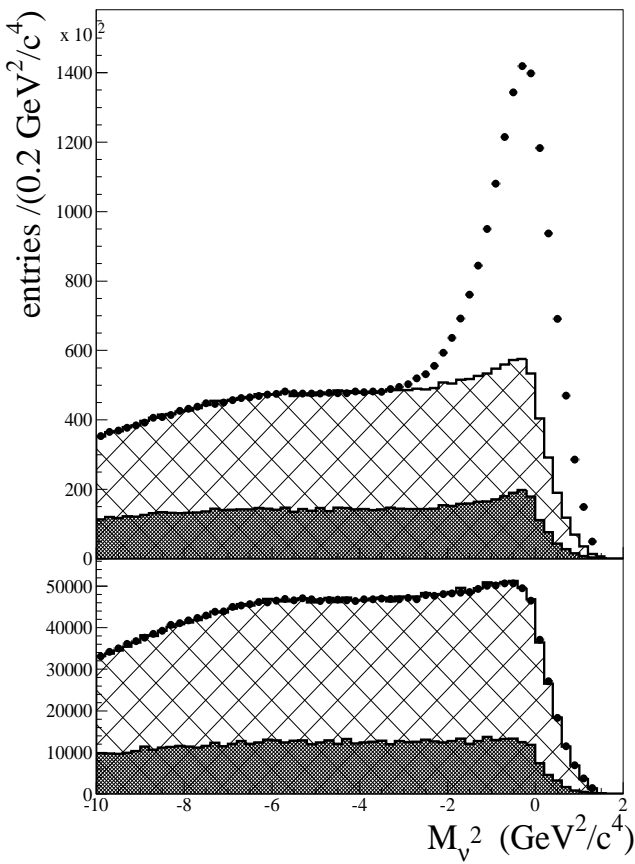}
\end{center}
\caption{$\mnusq$ distribution for right-charge (top) and wrong-charge (bottom) events.
The points correspond to on-resonance data. The distributions of continuum events 
(dark histogram), obtained from luminosity-rescaled off-resonance events, and \BB\ combinatorial background 
events (hatched area), 
obtained from the simulation, are overlaid. Monte Carlo events are normalized to the difference between 
on-peak and rescaled off peak data in the region $\mnusq<-4.5 $ GeV$^2/c^4$.
}
\label{f:mnu}
\end{figure}

To suppress leptons from several background sources, we use only high-momentum
leptons, in the range $ 1.3 < p_{\ell} < 2.4 $ GeV/$c$ \cite{ref:footnote2}.
The \psoft candidates have momenta ($p_\psoft$)
between 60 and 200 \mevc. Due to the limited phase space available in the \dsp\ decay,
the \psoft is emitted within an approximately one-radian half-opening-angle
cone centered about the \dsp\ flight direction.
We approximate the direction of the \dsp\ to be that of the \psoft
and estimate the energy  $\tilde{E}_{D^{*+}}$ of the \dsp\ 
as a function of the energy of the \psoft 
using a third order polynomial, with parameters taken from the simulation.
We define the square of the missing neutrino mass as
\ba
\label{e:m2}
\mnusq = \left( \frac{\sqrt{s}}{2} - \tilde{E}_{\dsp} - E_{\ell^-} \right)^2 - 
(\tilde{\bf{p}}_{\dsp} + {\bf{p}}_{\ell^-} )^2 ,
\ea
where we neglect the momentum of the \Bzb\ in the \FourS\ frame (on average, 0.34 GeV/$c$), 
and identify the \Bzb\ energy with the beam energy $\sqrt{s}/2$ in the \epem\ center-of-mass frame. 
$E_{\ell^-}$ and ${\bf{p}}_{\ell^-}$ are the energy and momentum vector of the lepton and 
$\tilde{\bf{p}}_{\dsp}$ is the estimated momentum vector of the \dsp. 
The distribution of $\mnusq$ peaks at zero for signal events, while it is spread over a
wide range for background events (see Fig.~\ref{f:mnu}).\par
We determine the \Bzb\ decay point from a vertex fit of the $\ell^-$ and \psoft tracks,
constrained to the beam-spot position in the plane perpendicular to the beam axis
(the $x$-$y$ plane). The beam-spot position and size are determined on a run-by-run basis using two-prong events
\cite{ref:babar}. Its size in the horizontal ($x$) direction is on average 120 $\mu$m. Although the beam-spot 
size in the vertical ($y$) direction is only 5.6 $\mu$m, we use a constraint of 50 $\mu$m in the vertex fit
to account for the flight distance of the \Bzb\ in the $x-y$ plane. We reject events
for which the $\chi^2$ probability of the vertex fit, ${\cal P}_V$, is less than $0.1\%$.\par
\begin{figure}[t]
\begin{center}
\includegraphics[width=9cm,height=11cm]{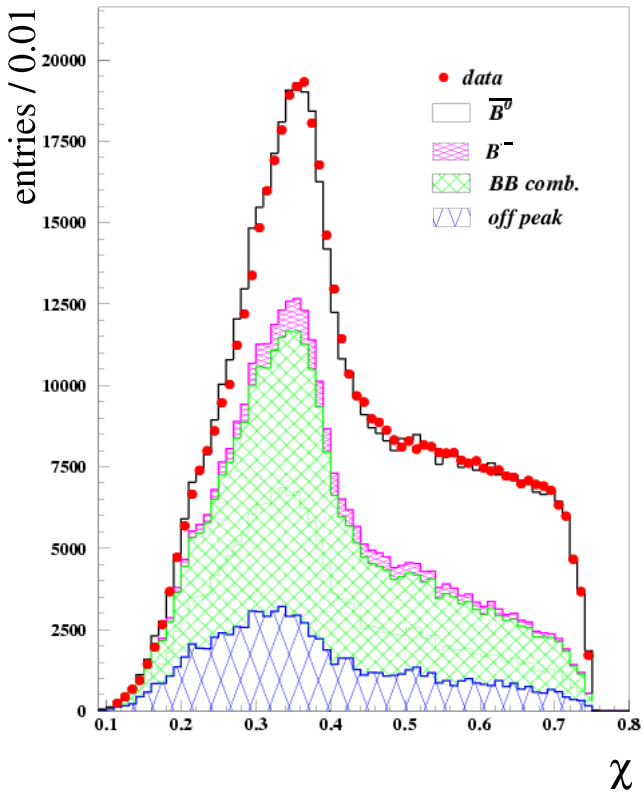}
\caption{ \label{f:X} Distribution of the combined signal likelihood ${\cal X}$
for events in the signal \mnusq\ region. Events for which ${\cal X}<0.4$ are rejected.}
\end{center}
\end{figure}

We then apply a selection criterion to a combined signal likelihood, ${\cal X}$, calculated 
from $p_{\ell^-}$, $p_\psoft$, and  
${\cal P}_V$, which results in a signal-to-background ratio of about one in the
signal region defined as $\mnusq > -2.5$ GeV$^2/c^4$. We reject events for which  ${\cal X}$ is lower than~$0.4$
(see Fig.~\ref{f:X}). 
Figure \ref{f:mnu} shows the distribution
of $\mnusq$ after this selection.
The distributions in the top part
of the figure are obtained from events in which the $\ell$ and the $\pi_s$ have opposite 
charges (``right-charge''), and the distributions in the bottom are from events in which the $\ell$ 
and the $\pi_s$ have equal charges (``wrong-charge''). 

  The points in Fig.~\ref{f:mnu} correspond to on-resonance data.
  The  dark histograms correspond to off-resonance data, scaled by the
  ratio of on-resonance to off-resonance integrated luminosity. 
  The hatched histograms correspond to \BB\ combinatorial background
  from simulation.
  To normalize the \BB\ combinatorial background, 
  we scale the \BB\ Monte Carlo histogram so that, when
  added to the luminosity-scaled off-resonance histogram, 
  the sum matches the on-resonance data in the  
  region $\mnusq < -4.5$ GeV$^2/c^4$.
  The right-charge plot is shown for illustration only.  
  We use the wrong-charge samples as a cross-check 
  to verify that the \BB\ combinatorial background shape is described by the simulation. For this
  purpose, we compare the number of wrong-charge events in the signal region 
  predicted from the sum of off-resonance
  and \BB\ Monte Carlo, normalized as above, to the number of wrong-charge 
  on-resonance data events.
  This ratio is $0.996 \pm 0.002$, consistent with unity. 
  For the rest of the analysis we consider only right-charge events.

\subsection{Tag Vertex and B Flavor Tagging}
 
To measure \deltamd\ we need to know the flavor of both $B$ mesons at their time 
of decay and their proper decay time difference $\deltat$. 
The flavor of the partially reconstructed $B$ is determined from the charge of the
high-momentum lepton. 
In order to identify the flavor of the other (``tag'') $B$ meson, 
we restrict the analysis to events in which another charged lepton (the ``tagging lepton'') is found.
To reduce contamination from fake leptons and leptons originating from charm decays, we require that 
the momentum of this second lepton exceed 1.0 GeV/$c$ for electrons, and 1.1 GeV/$c$ for muons.

The decay point of the tag $B$ is determined with the high-momentum lepton and a beam-spot constraint; 
the procedure is the same as that used to determine the $\overline{B}^0$ vertex.
We compute \deltat\ from the 
projected distance between the two vertices along the beam direction ($z$-axis), $\deltaz=z_{decay}-z_{tag}$,
with the approximation that the \Bz\ and the \Bzb\ are at rest in the \FourS\ rest frame (the boost approximation):
$\deltat = \deltaz/c \beta\gamma$,
where the boost factor $\beta\gamma$ is determined from the measured 
beam energies. 
To remove badly reconstructed vertices we reject all events with either $|\deltaz|>3$ mm 
or $\sigma(\deltaz) >0.5$ mm, where $\sigma(\deltaz)$ is the uncertainty on $\deltaz$,
computed for each event. 
The simulation shows that the difference between the true and measured  $\deltat$
can be fitted with the sum of two Gaussians. The rms of the narrow Gaussian, which
describes 70$\%$ of the events, is 0.64~ps; the rms of the wide one is about 1.7~ps.
 
We then select the best right-charge candidate in each event
according to the following procedure: if there is more than
one, we choose that with $\mnusq>-2.5$ Gev$^2$/c$^4$. If
two or more candidates are still left,
but they have different leptons, we
select the one with the largest value of $\chi$.
In a small fraction of events we select
two or more candidates sharing the same lepton
combined with different soft pions.
We keep the candidate with the largest $\chi$, unless
one of the \psoft\ is consistent with coming
from the decay of a \dsp\
from the other $B$, in which case we remove the event.
For this purpose, we define the square of the missing neutrino mass in the tag-side,
$\mnusq_{,tag\mathrm{-}B}$, by means of Eq.~\ref{e:m2}, where we replace the four-momentum
of the lepton from the \BtoDs\ decay with that of the tag lepton.
This variable peaks at zero for soft pions originating from the tag-$B$ decay. 
We require $\mnusq_{,tag\mathrm{-}B}<-3$ GeV$^2/c^4$. 
Finally we reject the events in which the signal lepton
can be combined to a wrong-charge pion to produce an
otherwise successful candidate, if the pion is consistent
with coming from a \dsp\ from the tag-$B$ according to the
criterion just described. About 20$\%$ of the signal events
are removed by this requirement.

For background studies, we select events in the region $\mnusq < -2.5~$GeV$^2/c^4$
if there is no candidate in the signal region. 
We find about 49000 signal events over a background of about 28000 events in the data sample
in the region $\mnusq > -2.5~$GeV$^2/c^4$.

\subsection{Sample Composition}
\label{s:sample}
Our data sample consists of the following event types, categorized according to their origin and 
to whether or not they peak in the $\mnusq$ distribution. 
We consider signal to be any combination of a lepton and a charged $D^*$ produced in the decay of a single 
\Bzb\ meson. Signal consists of mainly \BtoDs\ decays, with
minor contributions from $\Bzb \ra \dsp \pi^0 \ellm \nulb $,    
$\Bzb \ra \dsp \tau^- \nub_\tau $, $\Bzb \ra \dsp D_s^- $, and $\Bzb \ra \dsp \overline{D} X$ with $\tau$, 
${D_s}^-$, or $\overline{D}$
decaying to an $\ellm$, and from $\Bzb \ra \dsp h$, with the hadron $h$ misidentified as a muon.
Peaking \Bub\ background is mainly due to the processes $\Bub \ra \dsp \pi^- \ellm \nulb $, and
$B^- \ra D^{*+} \pi^- X$ with the $\pi^-$ misidentified as a muon. 
Other minor contributions to the peaking sample are due to decays $B \ra D^* \pi \nu_\tau \tau$ 
($\tau \ra \ell X$), $B \ra D^* \pi \overline{D} X $ ($\overline{D} \ra \ell Y$), 
where the $D^*$ and the $\pi$ come from the decay of an orbitally excited $D$ meson (\dstrstr). 
Non-peaking 
contributions are due to random combinations of a charged lepton candidate and a low-momentum pion candidate,
produced either in \BB\ events (\BB\ combinatorial) or in $e^+e^- \ra q\overline{q}$ interactions 
with $q=u,~d,~s$, or $c$
(continuum). We compute the sample composition separately for mixed and unmixed
events by fitting the corresponding $\mnusq$ distribution to the sum of four components: continuum, \BB\ 
combinatorial background, \BtoDs\ decays, and \BtoDss\ decays. Due to 
one or more additional pions in the final state, the \BtoDss\ events have a different \mnusq\ 
spectrum from that of the process \BtoDs.
 We measure the continuum contribution from the off-resonance sample,
scaled to the luminosity of the on-resonance sample. 
We determine the $\mnusq$ distributions for the other event types from the simulation, and determine 
their relative abundance in the selected sample from a fit to the $\mnusq$ distribution for the data.
Assuming isospin conservation, we assign two thirds of \BtoDss\ decays to peaking \Bub\ background and
the rest to 
$\Bzb \ra D^* \pi \ellm \nulb $
, which we add to the signal.
We vary this fraction in the study of systematic uncertainties.
We assume $50\%$ uncertainty on the isospin-conservation hypothesis. 
\par
A possible distortion in the $\mnusq$ distribution comes 
from the decay chain \mbox{$\overline{B} \ra D(X) \ellm \nulb$},~\mbox{$D\ra Y\pi^+$},
where the state $Y$ is so heavy that the charged pion is emitted at low momentum, behaving like a \psoft.
This possibility has been extensively studied by the CLEO collaboration~\cite{CLEODtokpi}, 
where the
three $D^+$ decay modes most likely to cause this distortion
have been identified: \mbox{$\overline{K}^{*0}\omega\pi^+$}, \mbox{$K^{*-}\rho^+\pi^+$}, 
and \mbox{$\overline{K}^{*0}\rho^0\pi^+$}. 
If we remove these events from the simulated sample, and we repeat the fit, 
the number of fitted signal events increases by 0.4\%. We assume therefore $\pm 0.4 \%$ systematic
error on the fraction of signal events in the sample due to this uncertainty.
\par
\begin{figure}[!htb]
\begin{center}
\begin{tabular}{c}
\hs{-.5cm}\includegraphics[width=9.4cm,height=9.4cm]{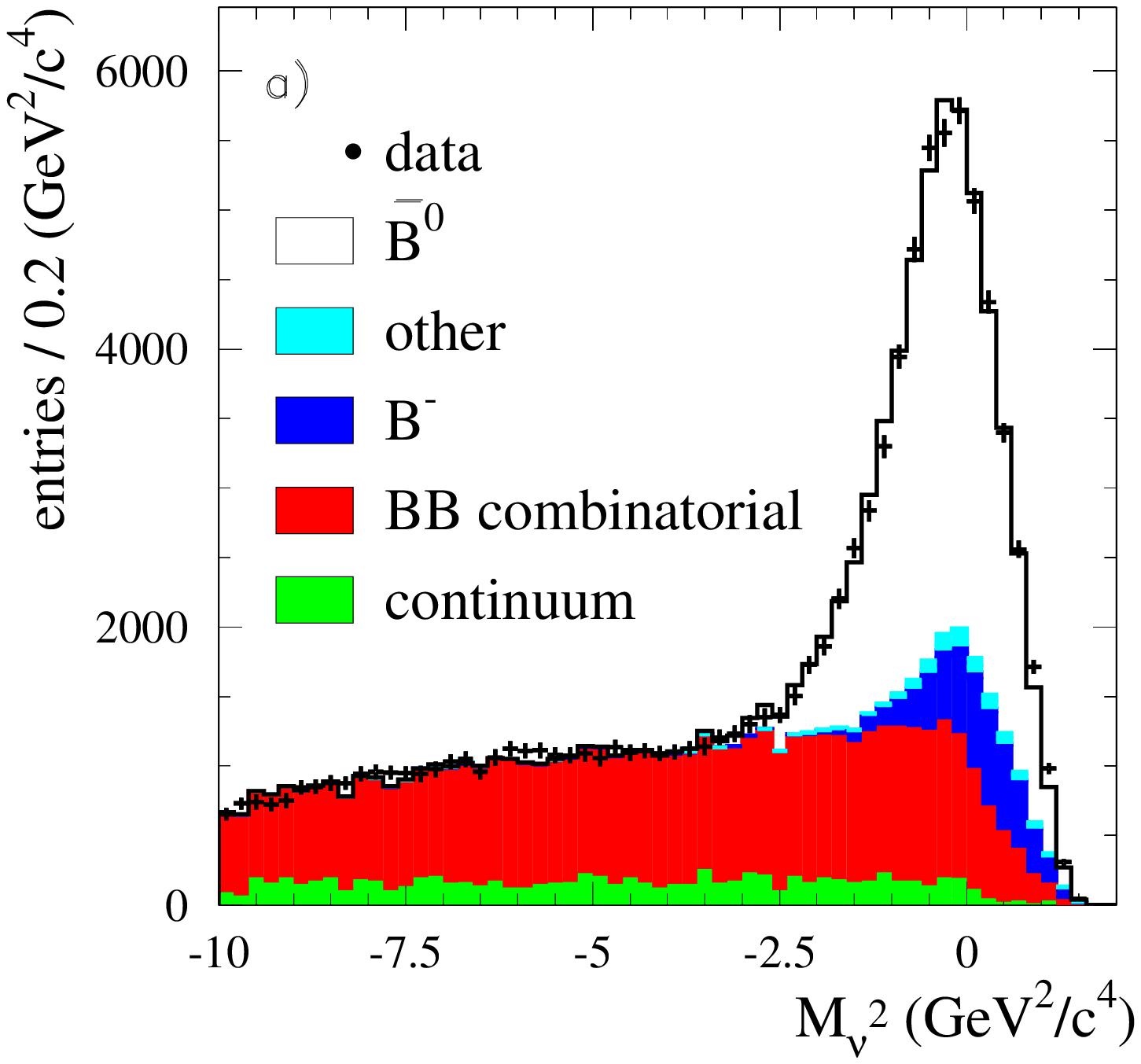}\\ 
\hs{-.5cm} \includegraphics[width=9.4cm,height=9.4cm]{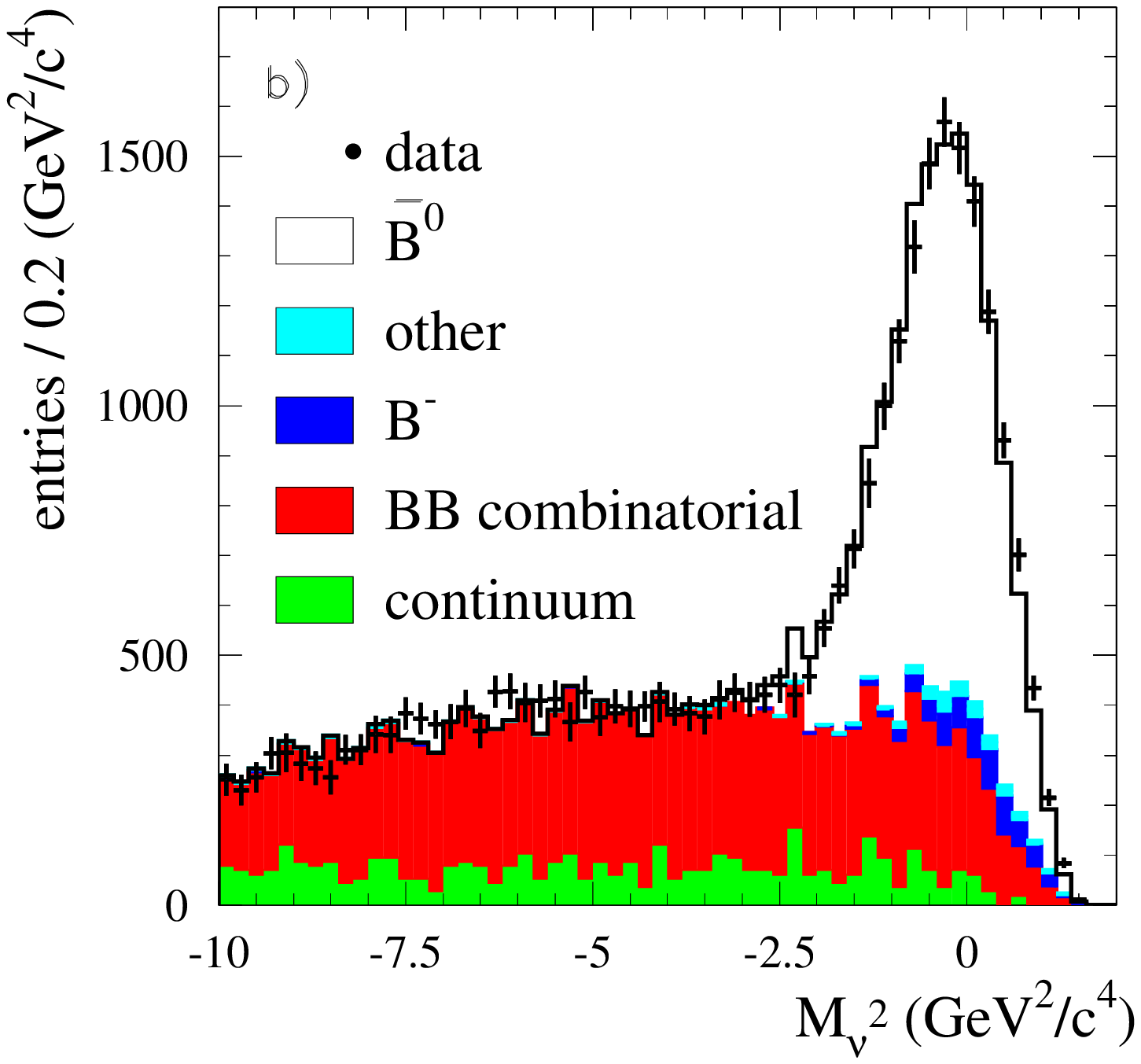}
\end{tabular} 
\end{center}
\caption{Fit to the $\mnusq$ distribution for the unmixed events 
(a) and mixed events (b).
``$\Bzb$'' includes $\BtoDs$, $\Bzb \ra \dsp \pi^0 \ell^- \bar{\nu}_\ell$,   
$\Bzb \ra \dsp \tau^- \bar{\nu}_\tau $ ($\tau \ra \ell X$), 
$\Bzb \ra \dsp D_s^- $ ($D_s \ra \ell X$), $\Bzb \ra \dsp \overline{D} X$ 
($\overline{D} \ra \ell Y$), 
and $\Bzb \ra \dsp h$ with the hadron $h$ misidentified as a muon.
``$B^-$'' includes $B^- \ra D^{*+} \pi^- \ell^- \bar{\nu}_\ell$ and $B^- \ra D^{*+} \pi^- X$ with the $\pi^-$ misidentified as a muon. ``Other'' includes $B \ra D^* \pi \nu_\tau \tau$  ($\tau \ra \ell X$), $B \ra D^*\pi \overline{D} X $ 
($\overline{D} \ra \ell Y$). 
 }
 \label{f:fitmnu}
\end{figure}
\begin{figure}[!htb]
\vs{-2.cm}
\begin{center}
\begin{tabular}{c}
\vs{-1.7cm}
\hs{-.5cm}\includegraphics[width=9cm,height=10.5cm]{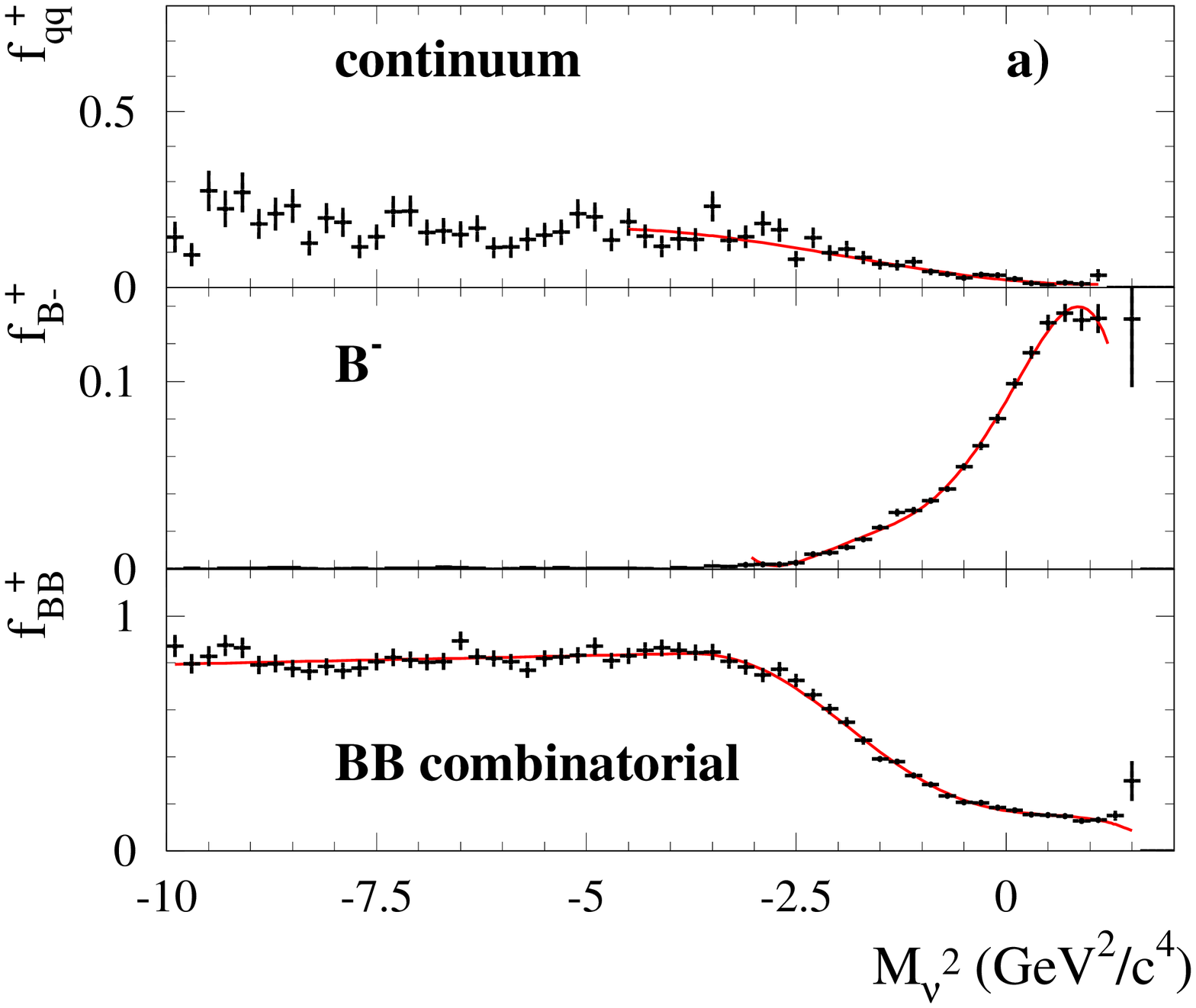}\\ 
\hs{-.5cm}\includegraphics[width=9cm,height=10.5cm]{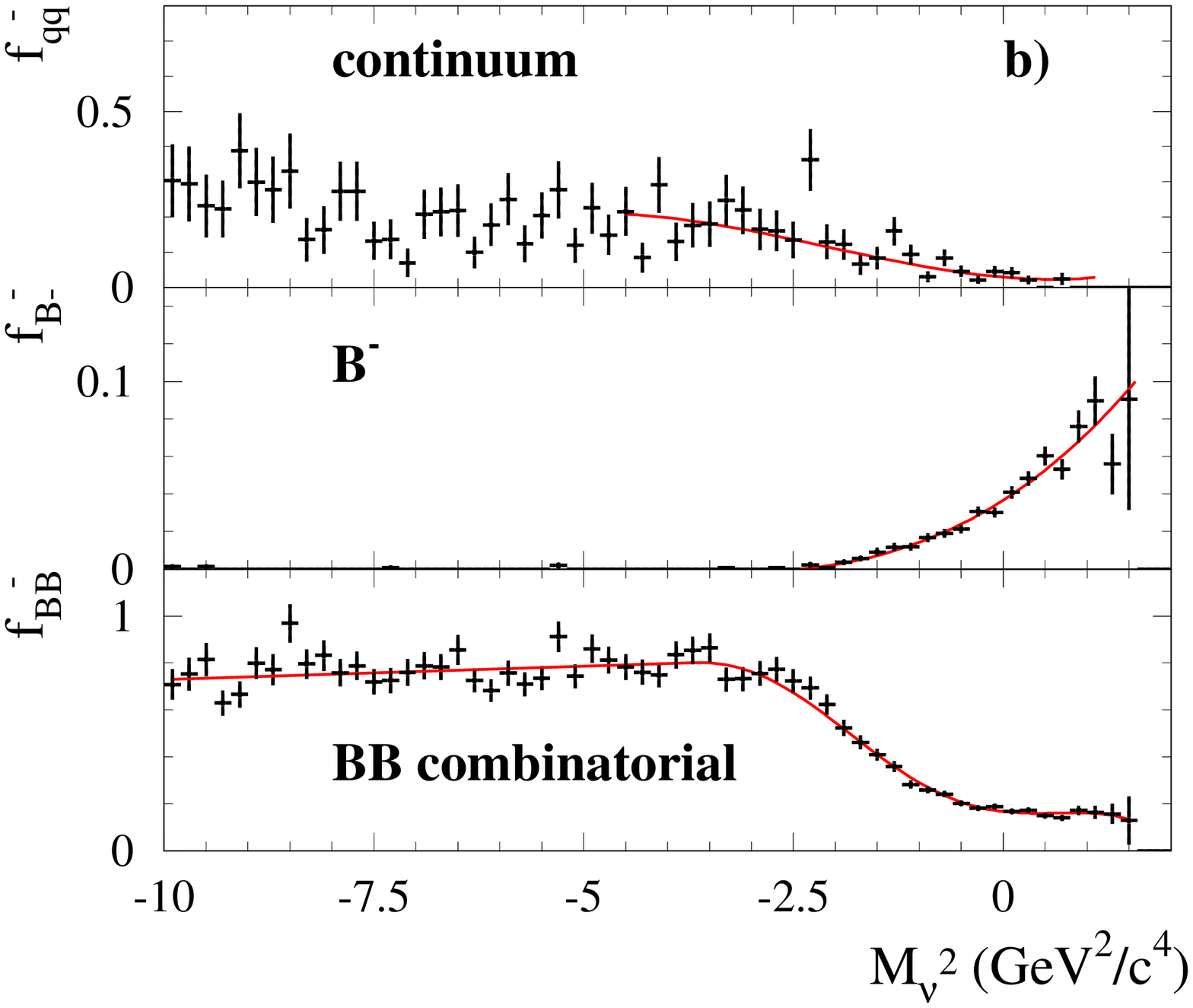}
\end{tabular} 
\end{center}
\caption{Fraction of continuum events, peaking $\Bub$, and \BB\ combinatorial background 
in the unmixed (a) and mixed (b) lepton-tagged samples. 
The continuous lines overlaid represent the analytic
functions ($f^{\pm}_{qq},f^{\pm}_{B^-}$ and
$f^{\pm}_{B \bar{B}}$) used to parameterize the distributions. 
The fraction of continuum events is parameterized only in the 
region $\mnusq>-4.5$~GeV$^2/c^4$. For $\mnusq<-4.5$~GeV$^2/c^4$, where just continuum and combinatorial 
backgrounds are present, we assume that $f_{\Bub}^{\pm}=0$, and we compute 
$f_{qq}^{\pm}=1-f_{\BB}^{\pm}$. 
}
\label{f:frac}
\end{figure}

Figure \ref{f:fitmnu} shows the \mnusq\ fit results for unmixed (upper) and mixed (lower) events.
We use the results of this study to determine the
fraction of continuum ($f_{qq}^{\pm}$), \BB\ combinatorial ($f_{\BB}^{\pm}$), and peaking \Bub\ 
($f_{\Bub}^{\pm}$) background
as a function of $\mnusq$, separately for mixed ($f^-$) and unmixed ($f^+$) events. We parameterize 
these fractions with polynomial functions of \mnusq\, as shown in Fig.~\ref{f:frac}.

\subsection{\boldmath \tBz and \deltamd Determination}
We fit data and Monte Carlo events with a binned maximum-likelihood method. 
We divide the events into one hundred
\deltat\ bins, spanning the range $-18$~ps $< \deltat < 18$~ps, and twenty \st\ bins between 0 and 3 ps.
We assign to all events in each bin the values of \deltat\ and \st\ corresponding to the
center of the bin. We fit simultaneously the mixed and unmixed events.
We maximize the likelihood

\ba
\label{e:likel}
{\cal L}=\left(\prod_{k=1}^{{\cal N}_{\mathrm{unmix}}}{\cal F}^{+}_{k}\right)
\left(\prod_{j=1}^{{\cal N}_{\mathrm{mix}}}{\cal F}^{-}_{j}\right)
\times C_{{\cal S}_{\Bub}}\times C_{\chi_d}, 
\ea
where the indices $k$ and $j$ denote the unmixed and mixed selected events.
The functions ${\cal F}^{\pm}(\deltat,\sigma_{\deltat},\mnusq | \tBz,\deltamd)$ 
describe the normalized \deltat\ distribution as the sum of the decay probabilities for signal 
and background events:
\ba
\nonumber
\label{e:SumPdf}
{\cal F}^{\pm}(\deltat,\sigma_{\deltat},\mnusq | \tBz,\deltamd) = \\
f^\pm_{qq}(\mnusq) \cdot {\cal F}^{\pm}_{qq}(\deltat,\sigma_{\deltat})\\  \nonumber + 
f^\pm_{\BB}(\mnusq) \cdot {\cal F}^{\pm}_{\BB}(\deltat,\sigma_{\deltat})\\ \nonumber +
{\cal S}_{\Bub}f^\pm_{\Bub}(\mnusq)  \cdot {\cal F}^{\pm}_{\Bub}(\deltat,\sigma_{\deltat})\\ \nonumber +
[1-{\cal S}_{\Bub}f^\pm_{\Bub}(\mnusq)-f^\pm_{\BB}(\mnusq)-f^\pm_{qq}(\mnusq)] \\ \nonumber \cdot
{\cal F}^{\pm}_{\Bzb}(\deltat,\sigma_{\deltat}|\tBz,\deltamd), 
\ea
where the functions ${{\cal F}_i^{\pm}}$ represent the probability density functions 
(PDF) for signal ($i=\Bzb$),
peaking \Bub ($i =\Bub $), \BB\ combinatorial ($i=\BB $), and continuum ($i=q\overline{q}$) events, 
modified to account for the finite 
resolution of the detector, 
and the superscript +($-$) applies to unmixed (mixed) events.
The resolution function is expressed as the sum of three Gaussian 
functions, described 
as ``narrow'', ``wide'', and ``outlier'':
\ba
\nonumber
{\cal R}(\delta\deltat,\sigma_{\deltat}) &=& 
  \frac{(1-f_w-f_o)}{\sqrt{2\pi} S_{n} \sigma_{\deltat}} ~
e^{-\frac{(\delta\deltat - o_{n})^2}{2 S_{n}^2 \sigma_{\deltat}^2}}  \\ \nonumber &+&
 \frac{f_w }{\sqrt{2\pi} S_{w} \sigma_{\deltat}} ~
e^{-\frac{(\delta\deltat - o_{w})^2}{2 S_{w}^2 \sigma_{\deltat}^2}}  \\ \nonumber &+&
 \frac{f_o}{\sqrt{2\pi} S_o}  ~
 e^{-\frac{(\delta\deltat - o_{o})^2}{2 S_{o}^2 }} ,
\ea
 where $\delta\deltat$ 
is the difference between the measured and true values of \deltat , $o_{n}$ and $o_{w}$ 
are offsets, and the factors $S_{n}$ and $S_{w}$
account for possible misestimation of $\sigma_{\deltat}$. The outlier term, described by 
a Gaussian function of fixed width $S_{o}$ and offset $o_{o}$, is introduced to
describe events with badly measured $\Delta t$, and accounts for less than 1$\%$ of the events.
\par
To account for the $\pm 50\%$ uncertainty on the isospin
assumption (see Sec. \ref{s:sample}), 
the functions $f^+_{B^-}$ and $f^-_{B^-}$ are multiplied in
the PDF for the peaking $B^-$ background
by a common scale factor ${\cal S}_{B^-}$. This parameter
is allowed to vary in the fit, constrained to unity with variance $\sigma_{\Bub}=0.5$ 
by means of the Gaussian term 
\ba
\nonumber
C_{{\cal S}_{\Bub}}=e^{-\frac{({\cal S}_{\Bub}-1)^2}{2 \sigma_{\Bub}^2}}.
\ea
\par
We constrain the expected fraction $P_{\mathrm{exp}}$ of mixed events to the observed one 
\ba
\nonumber
P_{\mathrm{obs}}=\frac{{\cal N}_{\mathrm{mix}}}{{\cal N}_{\mathrm{mix}}+{{\cal N}_{\mathrm{unmix}}}}
\ea
by means of the binomial factor 
\ba
\nonumber
C_{\chi_d}=\frac{{\cal N}!}{{\cal N}_{\mathrm{mix}}!{\cal N}_{\mathrm{unmix}}!}P_{\mathrm{exp}}^{{\cal N}_{\mathrm{mix}}}(1-P_{\mathrm{exp}})^{{\cal N}_{\mathrm{unmix}}}.
\ea
For a sample of signal events with dilution ${\cal D}$, the expected fraction reads
\ba
\nonumber
P_{\mathrm{exp}}(\dmd, \tBz, {\cal D})= \chid \cdot {\cal D} + \frac{1-{\cal D}}{2},
\ea
where, neglecting the decay-rate difference $\Delta \Gamma_d$ between the two mass eigenstates,
 the integrated mixing rate \chid\ is related to the product $x = \dmd \cdot \tBz$  by the relation
\ba
\nonumber
\chid = \frac{x^2}{2(1+x^2)}.
\ea 
\par

We divide signal events according to the origin of the tag lepton into primary ($\ptl$), cascade ($\ctl$), 
and decay-side ($\cdl$) lepton tags.
A primary lepton tag is produced in the direct decay $\Bz \ra \ell^+ \nu_{\ell} X$. 
These events are described by 
Eq.~\ref{eq:pdf}, with ${\cal D}$ close to 1 (a small deviation from unity is expected due to hadron misidentification,
leptons from $J/\psi$, etc.). We expect small values of $o_n$ and $o_w$ for primary tags, 
because the lepton originates from the \Bz\ decay point.\par
Cascade lepton tags, produced in the process $\Bz \ra D X, D \ra \ell Y$, are suppressed by the requirement on the 
lepton momentum but still exist at a level of $9\%$, which we determine by varying their relative 
abundance $(f_{\ctl})$ as an additional parameter in the \deltamd and \tBz fit on data.
The cascade lepton production point is displaced from the 
\Bz\ decay point due to the finite lifetime of 
charm mesons and the \epem\ energy asymmetry. This results in a significant negative value of the offsets 
for this category.
Compared with the primary lepton tag, the cascade lepton is more likely to have the opposite charge correlation with the \Bz\ flavor. The same charge correlation is obtained when the charm 
meson is produced from the hadronization of the virtual $W$ from \Bz\ decay, which can
result in the production of two opposite-flavor charm mesons. We account
for these facts by applying Eq.~\ref{eq:pdf} to the cascade tag events with negative 
dilution ${\cal D_{\ctl}} = -(1-2f_{\ctl}^{\overline{b}\ra c\ra \ell^+}) = -0.65 \pm 0.08$, where we take from the PDG \cite{ref:PDG}
the ratio: 
{\small
\ba
\nonumber
 f_{\ctl}^{\overline{b}\ra c\ra \ell^+} = 
\frac{{\cal B}(\overline{b}\ra c\ra \ell^+)}{{\cal B}(\overline{b}\ra c\ra \ell^+)+{\cal B}(\overline{b}\ra \overline{c}\ra \ell^-)} 
= 0.17\pm0.04 . 
\ea
}
The contribution to the dilution from other sources associated with the
$\psoft \ell^-$ candidate, such as fake hadrons, is
negligible.\par
\begin{figure}[!htb]
\begin{center}
\begin{tabular}{c}
\hs{-.5cm}\includegraphics[width=8.6cm,height=8.6cm]{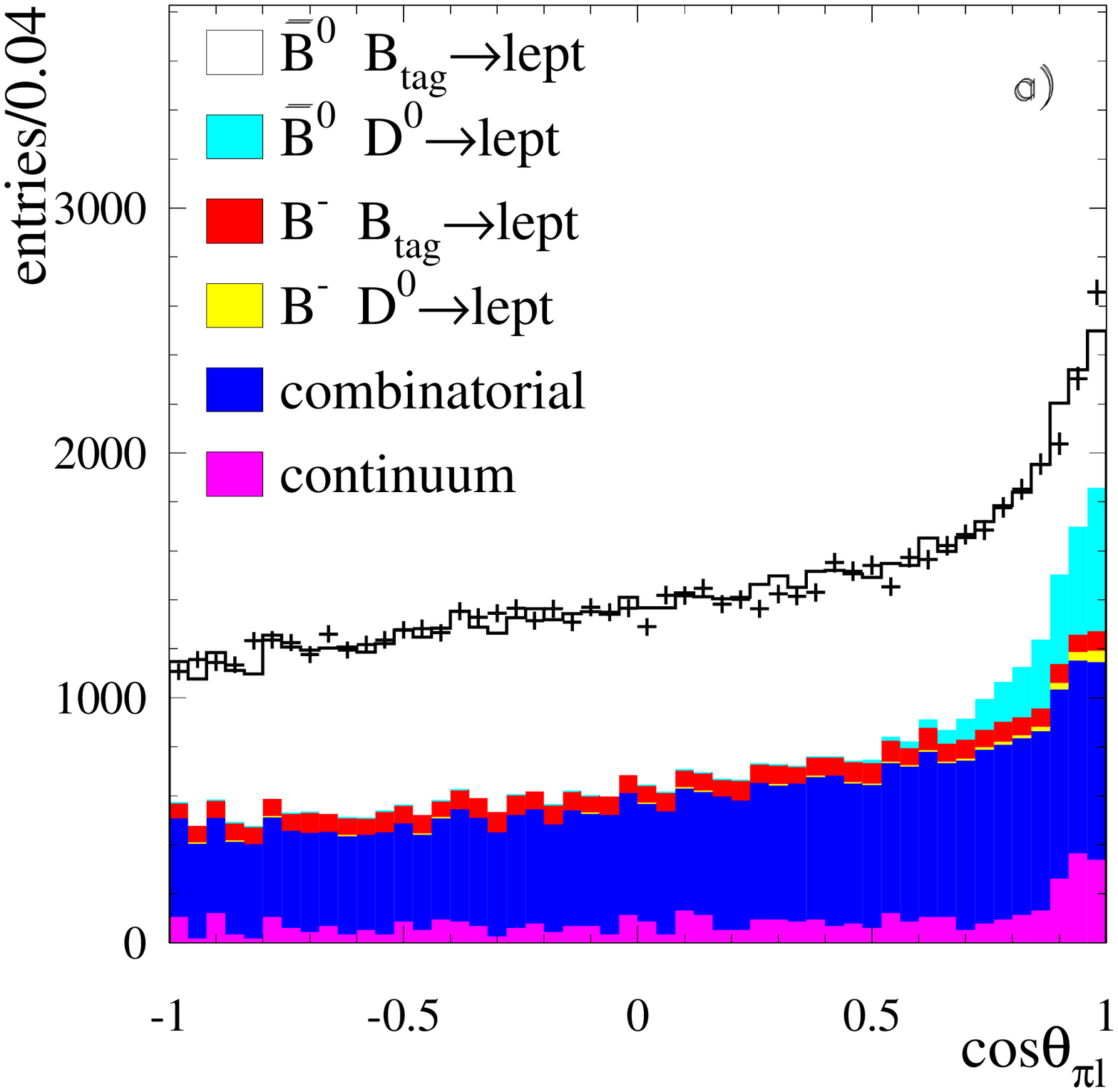}\\
\hs{-.5cm}\includegraphics[width=8.6cm,height=8.6cm]{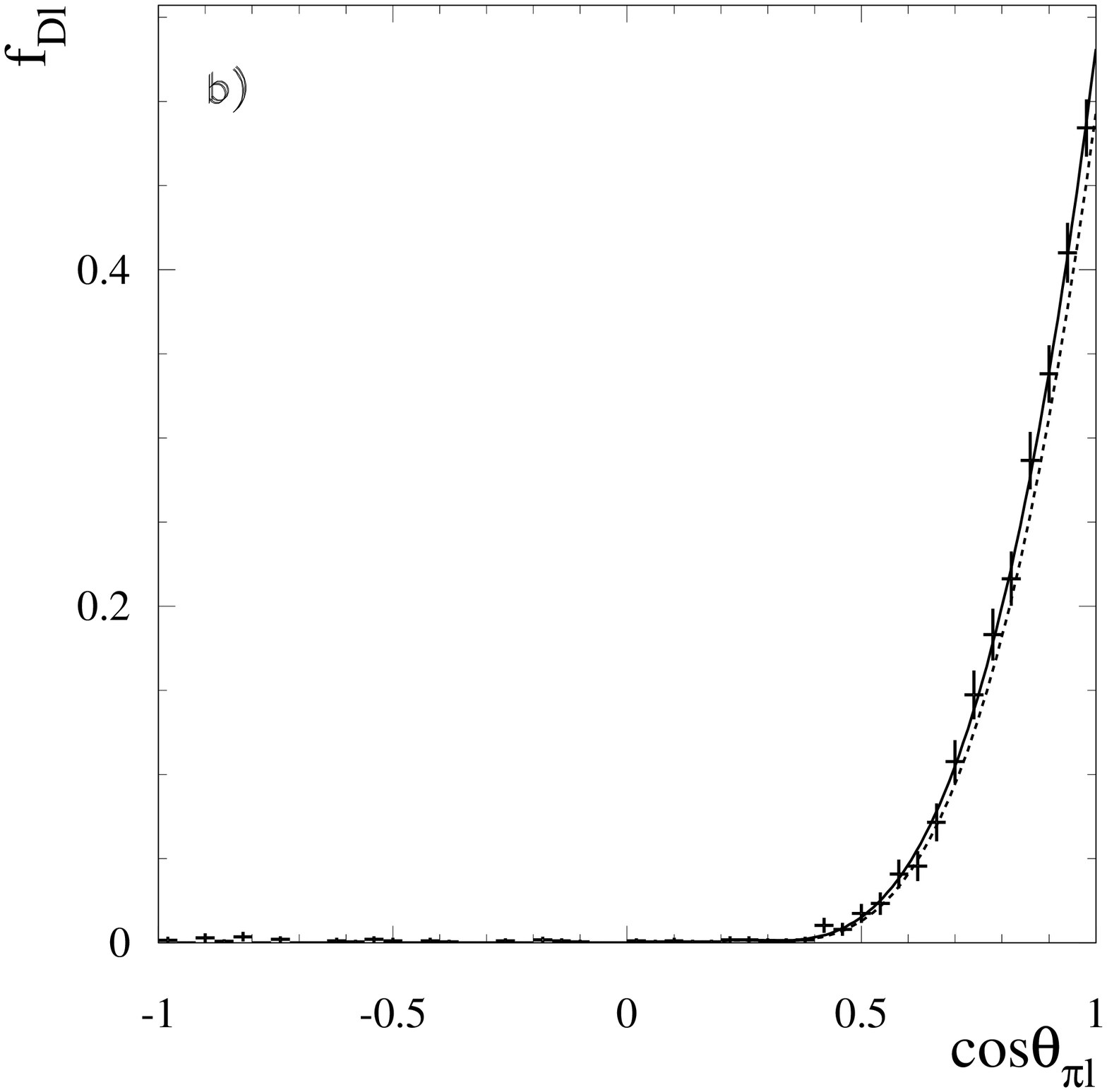}
\end{tabular} 
\end{center}
\caption{ Distribution of $\cos\theta_{\psoft \ell^-}$ for unmixed 
events. In figure (a), the points with error bars represent the data, and the histograms
show the various sample components determined by the fit. 
Figure (b) shows the ratio between the fraction of
tags from \Dz\ decays over the total number of tags in
signal events, as obtained from the simulation rescaled to
the result of the fit. This distribution is parametrized by
a third order polynomial represented by the continuous line overlaid.
The dotted line represents the corresponding distribution obtained
without rescaling the simulation to the fit result.
}
\vs{1.cm}
\label{f:costh1}
\end{figure}
\begin{figure}[!htb]
\begin{center}
\begin{tabular}{c}
\hs{-.5cm}\includegraphics[width=9cm,height=9cm]{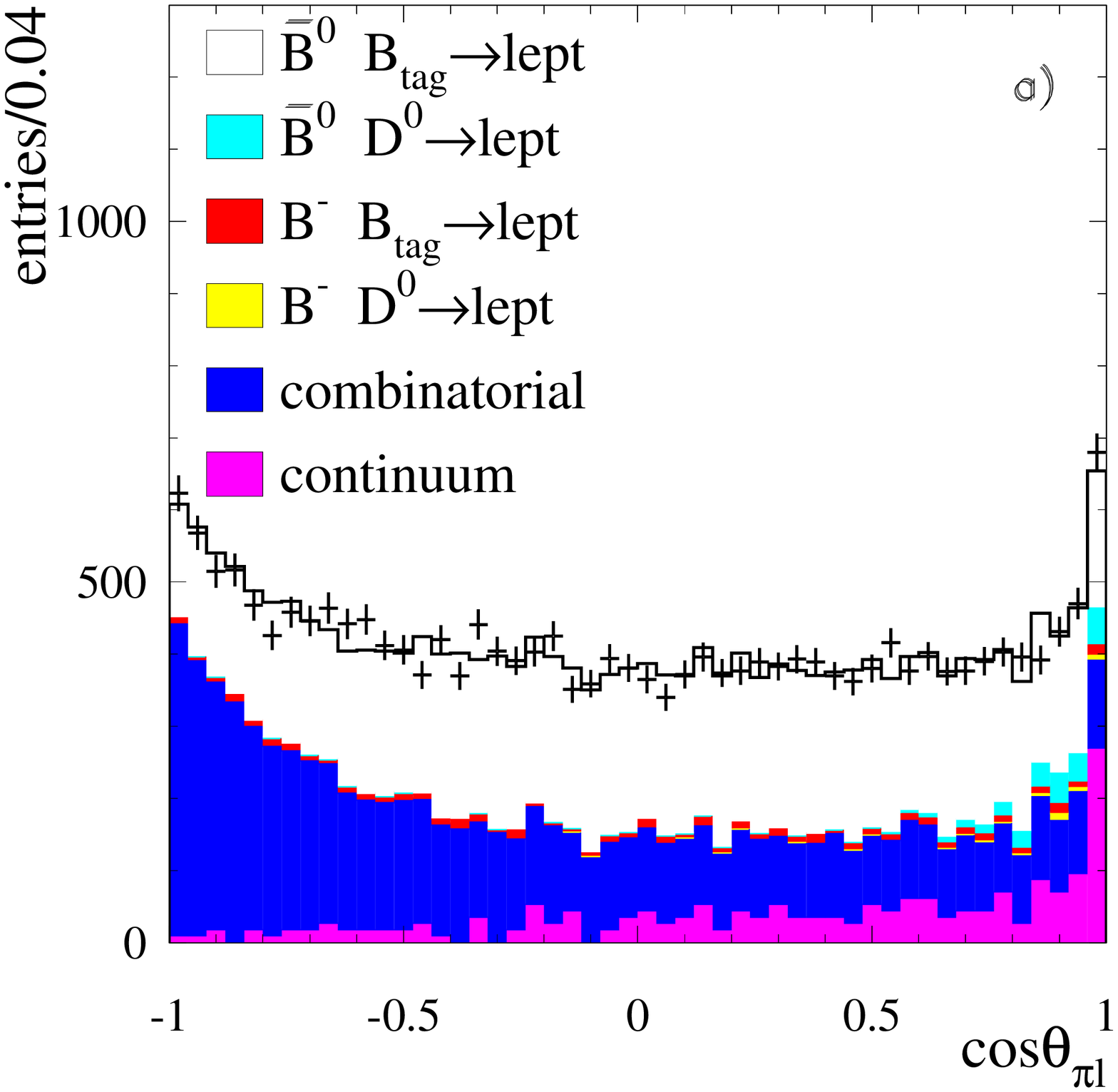}\\
\hs{-.5cm}\includegraphics[width=9cm,height=9cm]{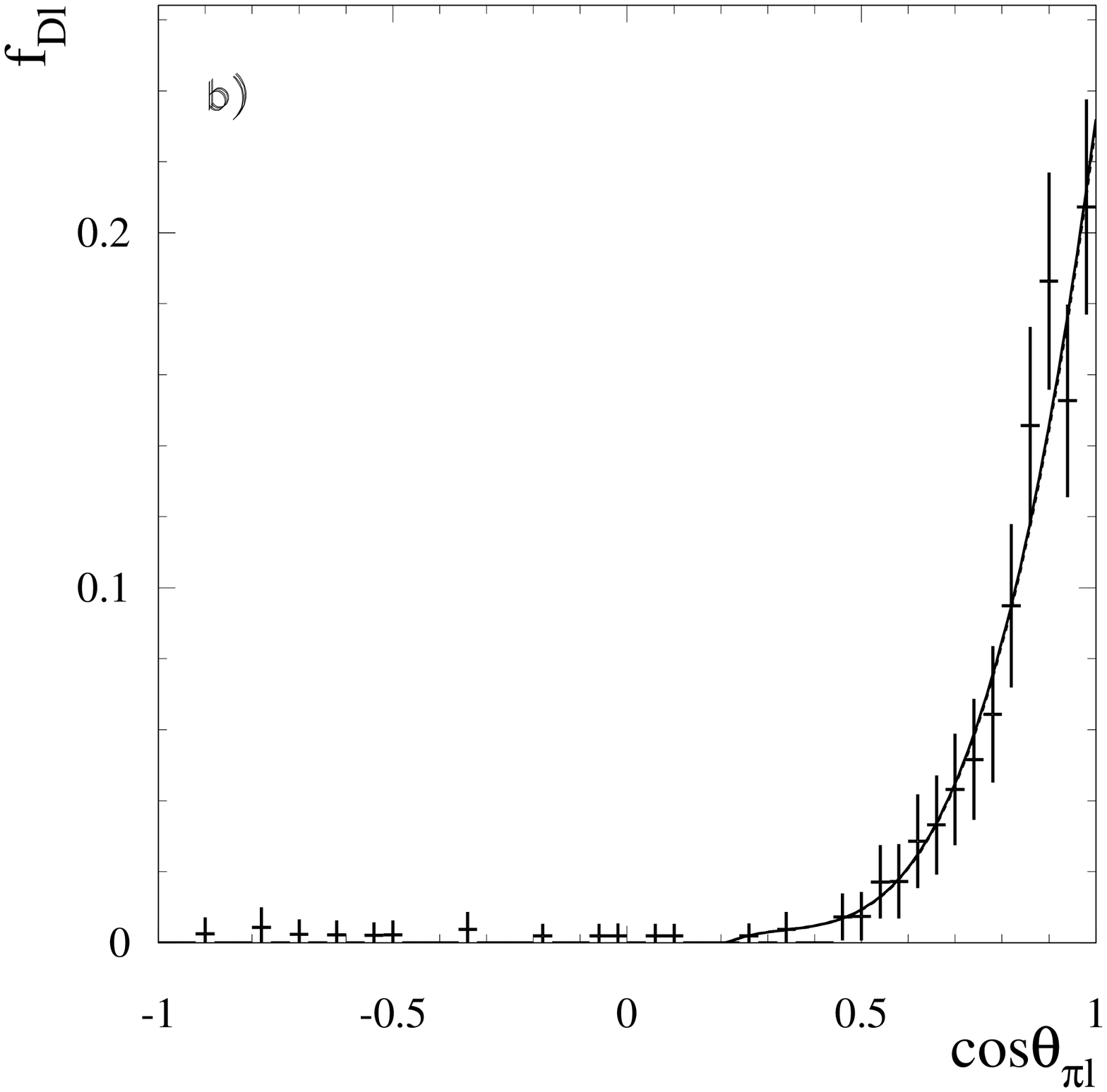}
\end{tabular} 
\end{center}
\caption{Same as Fig.~\ref{f:costh1} for mixed events.}
\label{f:costh2}
\end{figure}

Decay-side tags are produced by the semileptonic decay of the unreconstructed \Dz . Therefore
they do not carry any information about $\tBz$ or \deltamd.  The PDF for both mixed and unmixed contributions
is a purely exponential function, with an effective lifetime $\tDe$ representing the displacement of 
the lepton production point from the \Bzb\ decay point due to the finite lifetime of the \Dz .
We determine the fraction of these events by fitting 
the $\cos\theta_{\psoft \ell^-}$ distribution (see plots in Figs.~\ref{f:costh1}~(a) and \ref{f:costh2}~(a)),
where $\theta_{\psoft \ell^-}$ is the angle between the soft pion and the tag lepton in the $\epem$ 
rest frame. 
We fit the data with the sum of the histograms
for signal events, $\BB$ combinatorial background,
and peaking $\Bub$ background obtained from the simulation, 
and continuum background obtained from the off-resonance events. 
We fix the fraction of signal events, peaking $\Bub$ background, $\BB$ combinatorial background and continuum 
background in the fit and we allow to vary the relative amount of decay-side tags and tag-side tags.

Using the results of the  $\cos\theta_{\psoft \ell^-}$ fit we parameterize the probability for each event to have 
a decay-side tag as a third-order polynomial 
function of $\cos\theta_{\psoft \ell^-}$~(see plots in Figs.~\ref{f:costh1}~(b) and \ref{f:costh2}~(b)). 
\par The signal PDF for both mixed and unmixed events consists of the sum of PDFs for
primary, cascade, and decay-side tags,
each convoluted with its own resolution function. The parameters $S_n,~S_w,~S_o,~f_w$, and 
$f_o$ are common to the three terms, but each tag type has different offsets ($o_n,o_w$).
All the parameters of the resolution functions, the dilution of the primary tags, 
the fraction of cascade tags, and the effective lifetime of the decay-side tags are free parameters
in the fit. We fix the
other parameters (dilution of cascade tags, fraction of decay-side tags) to the values obtained
as described above, and then vary 
them within their uncertainties to assess the corresponding systematic error.
\par 
We adopt a similar PDF for peaking \Bub\ background, with separate primary, cascade, and decay-side terms.
Because \Bub\ mesons do not oscillate, we use a pure exponential PDF for the primary and cascade tags
with lifetime $\tBu = 1.671$~ps \cite{ref:HFAG}. We force the parameters of the resolution function
to equal those for the corresponding signal term. 
\par 
We describe continuum events with an exponential function convoluted with a three-Gaussian resolution
function. The mixed and
unmixed terms have 
a common effective lifetime $\tau_{lq}$. 
All the parameters of the continuum resolution function
 are set equal to those of the signal, except for the offsets, which are free in the fit.
\par
The PDF for combinatorial \BB\ background accounts for oscillating and non-oscillating subsamples.
It has the same functional form as the PDF for peaking events, but with
independent parameters for the oscillation frequency, the lifetimes and the fractions of 
$\Bub$ background, primary, cascade, and decay-side tag events. 
The parameters  $S_w$, $f_w$ and $o_o$ are set to the same values as those in the signal PDF.
\par

\section{RESULTS}
\label{sec:Physics}
We first apply the measurement procedure on several Monte Carlo samples.
We validate each term of the PDF by first fitting signal events, for
primary, cascade and decay-side tags separately, and then adding them together. We then
add peaking \Bub\ background, and finally add the \BB\ combinatorial background. We observe
the following features: 
\bi
\item The event selection introduces no bias on $\tBz$ and a bias of 
$(-0.0029\pm0.0010)$~ps$^{-1}$ on \deltamd.
\item The boost approximation introduces a bias on $\tBz$ ($+0.0054$~ps) and an additional bias on 
\deltamd\ ($- 0.0034$~ps$^{-1}$),
determined by fitting the true \deltaz\ distribution. These biases disappear however
when we fit the smeared \deltaz\ and allow for the experimental resolution in the 
fit function.
\item After the introduction of \Bub peaking background we observe a  
bias of $(-0.0079\pm0.0064)$ \ps\ on $\tBz$ and $(-0.0034\pm0.0028)$~ps$^{-1}$ on $\dmd$.
\item Adding combinatorial \BB\ events, we observe a bias of $(+0.0063\pm0.0070)$ \ps\ on $\tBz$
and $(-0.0074\pm0.0035)$~ps$^{-1}$ on \deltamd.
\item The isospin scale factor ${\cal S}_{\Bub} = 0.91 \pm 0.10$ 
is consistent with unity.
\ei
Based on these observations, we correct the data results by subtracting $0.0063$~ps
from $\tBz$, and adding $0.0074$~ps$^{-1}$ to $\dmd$.
We include the Monte Carlo statistical errors of $\pm0.0070$~ps for $\tBz$ and $\pm0.0035$~ps$^{-1}$
for $\dmd$ as systematic uncertainties.
\par
We determine the parameters for continuum events directly from the fit to on-resonance data,
and we independently fit the off-resonance events to verify the consistency with the on-resonance continuum
results.\par
We finally perform the fit to the on-resonance data. 
Together with \dmd\ and $\tBz$, 
we allow to vary 
most of the parameters describing the peaking \Bub , \BB\ combinatorial, and continuum background events.
The results of the fits to the Monte Carlo and data samples are shown in 
Tables~\ref{t:result1} and ~\ref{t:result2}.

The fit results are
$\tBz = (1.510 \pm 0.013 ~\mathrm{(stat))~ps,}$ and
$\dmd = (0.5035 \pm 0.0068 ~\mathrm{(stat))~ps}^{-1}$.
We correct these values for the biases measured in the Monte Carlo simulation,
obtaining the results
\ba
\nonumber
\label{e:res}
\tBz &=& (1.504 \pm 0.013 ~\mathrm{(stat))~ps,}\\
\nonumber \dmd &=& (0.5109 \pm 0.0068 ~\mathrm{(stat))~ps}^{-1}.
\ea
The statistical correlation between \deltamd\ and $\tBz$ is $0.7\%$. 
\deltamd\ has sizable correlations with
${\cal S}_{\Bub}$ (50$\%$) and with the fraction of cascade tags (24$\%$). $\tBz$ is correlated 
with $S_n$ ($-27\%$) and the offset of the wide Gaussian for the cascade tags ($-31\%$).
The complete set of fit parameters is reported in Tables~\ref{t:result1} and \ref{t:result2}.
\begin{table*}[tbp]
\caption{\label{t:result1}Parameters used in the PDFs.
The upper set of parameters refers to peaking events; the lower one refers to those 
parameters of the resolution function that are common to all the event types.
The second column shows how the parameters are treated in the fit. 
The third (fourth) column gives the result of the fit on data (MC) for free parameters and
the value employed for the 
parameters that are fixed or used as a constraint. The quoted error is the statistical uncertainty from the
fit for free parameters and the range of variation used in the systematic error determination for the 
others.
The last column shows the sample in which the parameter is used.
$\ptl$, $\ctl$ and $\cdl$ refer to primary, cascade and decay-side lepton tags, respectively. 
The parameters $o$, $S$, and $f$ correspond to offsets, scale factors, and fractions in the 
resolution function.
}
\bc
{\small 
\begin{tabular}{|l|c|c|c|c|}
\hline
Parameter &       Usage      &  data           &  MC   &  Sample         \\ \hline \hline
\vs{-.3cm}
     &               &                    &                       & \\
$\tBz$ (ps)       &      free     &  1.510$\pm$0.013    & 1.554$\pm$0.007        &  \Bzb \ptl, \ctl \\
\dmd   (ps$^{-1})$ &      free     &  0.503$\pm$0.007    & 0.465$\pm$0.004        &    \Bzb \ptl, \ctl \\ 
\tDe   (ps)        &      free     &  0.12$\pm$0.04      & 0.21 $\pm$0.02        &    \Bzb, \Bub, \\
      & & &  &   \BB \cdl \\
$\tBu$ (ps)      &       fixed      & 1.671$\pm$0.018      & 1.65  & \Bub, \BB   \\
${\cal S}_{\Bub}$  & constr. & 1.05$\pm$0.15  & 0.91$\pm$0.10          &                   \\
                & (1.0$\pm$0.5) &                &                        &                   \\           
$f_{\ctl}$    &   free    &  0.095$\pm$0.006 &  0.066$\pm$0.004       &    \Bzb           \\
$D_{\ptl}$     &  free     &  1.00$\pm$0.02  &  0.970$\pm$0.006       &    \Bzb,\ptl      \\
$D_{\ctl}$  &    fixed &  -0.65$\pm$0.08   & -0.545 & \Bzb,\ctl \\
$o_{\ptl,n}$ &      free     & $-$0.019$\pm$0.011   &  $-$0.012$\pm$0.007  &    \Bzb, \Bub \ptl \\
$(=o_{\ptl,w})$   &     &    &  &  \\
$o_{\ctl,n}$      &      free     & $-$0.18$\pm$0.07     &  $-$0.43$\pm$0.07       &    \Bzb \ctl \\
$o_{\ctl,w}$       &      free     &  2.8$\pm$1.1     &  $-$5.8$\pm$0.7       &    \Bzb \ctl \\
$o_{\cdl,n}$ &   free  &  $-$0.12$\pm$0.03   &  $-$0.14$\pm$0.02       &    \Bzb, \Bub, \cdl \\
$(=o_{\cdl,w}$)    &     &     &       &   \\
$S_n$          &      free     &       0.952$\pm$0.015     &  1.007$\pm$0.006       &    all  \\
$S_o$ (ps)         &      free     &  12.8$\pm$5.6      &  17.9$\pm$8.2       &    \Bzb, \Bub \\
$f_o$           &      free     & 0.0013$\pm$0.0005 &  0.0008$\pm$0.0003       &    \Bzb, \Bub \\
\hline 
$S_w$           &      free     &       2.57$\pm$0.13       &  2.63$\pm$0.15       &    all  \\
$f_w$           &      free     &  0.050$\pm$0.005   &  0.035$\pm$0.005       &    all  \\
$o_o$           &       fixed      &   0     &    0    &   all  \\ \hline \hline
\end{tabular} }
\ec
\end{table*}

\begin{table*}[tbp]
\caption{\label{t:result2}
Parameters used in the background PDF.
The upper set of parameters refers to \BB\ combinatorial events, the central one refers to continuum 
parameters, and 
the lower set refers to those 
parameters of the resolution function that are common to all event types.
The symbols $\alpha$ correspond to the fractions of decay-side tags in the different samples. 
The last line shows the statistical correlation between \tBz and \dmd.
} 
\bc
{\small 
\begin{tabular}{|l|c|c|c|c|}
\hline
Parameter &       Usage      &  data           &  M.C.   &  Sample        \\
\hline \hline
\vs{-.3cm}
     &               &                    &                       & \\
$\tau_{B^0}^{BKG}$ (ps)  &      free     &  1.22$\pm$0.07    &  1.37$\pm$0.07       &  \BB \\
$\Delta m_d^{BKG}$ (ps$^{-1}$)  &      free     &  0.37$\pm$0.06    &  0.42$\pm$0.04       &  \BB \\ 
$\tBu^{BKG}$ (ps) &       fixed      & 1.671$\pm$0.018    & 1.65  & \Bub, \BB   \\
$f_{\ctl, u}^{BKG}$        &      fixed       &  0.030$\pm$0.006  &    0.030    &  \BB (\Bzb only) \\
$f_{\ctl, m}^{BKG}$   &      free     &  0.041$\pm$0.022  &    0.069$\pm$0.021  &  \BB (\Bzb only) \\
$f_{B-, u}^{BKG}$    &      free     &  0.62$\pm$0.08  &    0.52$\pm$0.02    &  \BB \\
$f_{B-, m}^{BKG}$    &      free     &  0.15$\pm$0.10  &    0.11$\pm$0.04    &  \BB \\
$\alpha_{B0, u}^{BKG}$  &   free     &  0.11$\pm$0.19  &  0.25$\pm$0.03       &  \BB (\Bzb only) \\
$\alpha_{B0, m}^{BKG}$ &   fixed     &  0.065$\pm$0.013   &  0.065      &  \BB (\Bzb only) \\
$\alpha_{B-, u}^{BKG}$  &   free     &  0.21$\pm$0.10  &  0.20$\pm$0.02       &  \BB (\Bub only) \\
$\alpha_{B-, m}^{BKG}$ &   fixed     &  0.36$\pm$0.07  &  0.36       &  \BB (\Bub only) \\
$f_{\cdl,2}$ &   fixed     &  0.60$\pm$.12  &  0.60$\pm$.12   &  \BB, \cdl  \\
$D_{\ptl}^{BKG}$     &      free     &  0.989$\pm$0.013  &    0.964$\pm$0.006    &  \BB  (\Bzb\ only)       \\
$D_{\ctl}^{BKG}$     &      fixed    & -0.65$\pm$0.08  &   -0.545   &  \BB  (\Bzb\ only)       \\
$o_{\ptl}^{BKG}$     &      free     &  $-$0.006$\pm$0.016    &   $-$0.02$\pm$0.03   &  \BB, \ptl \\
$o_{\ctl}^{BKG}$     &      free     &  $-$1.6$\pm$0.6      &  $-$0.8$\pm$0.2       &  \BB, \ctl \\
$o_{\cdl}^{BKG}$     &      free     &  $-$0.02$\pm$0.04   &  $-$0.05$\pm$0.03       &  \BB, \cdl  \\
$S_n^{BKG}$      &      free     &       0.961$\pm$0.015     &  0.961$\pm$0.021       &    all  \\
$S_o^{BKG}$ (ps)  &      free     &  10.4$\pm$3.6     &  14.7$\pm$6.1      & \BB  \\
$f_o^{BKG}$       &      free     &  0.0021$\pm$0.0009 &  0.0008$\pm$0.0003       & \BB  \\ \hline 
$\tau_{lq}$ (ps)     &      free     &  0.27$\pm$0.05    &    -     &  Continuum \\
$o_{lq,n}$  &      free     &  0.007$\pm$0.032   &    -     &  Continuum \\ 
$(=o_{lq,w})$  &      &   &        &  \\ \hline 
$S_w$           &      free     &       2.57$\pm$0.13       &  2.63$\pm$0.15       &    all  \\
$f_w$           &      free     &  0.050$\pm$0.005   &  0.035$\pm$0.005       &   all  \\
$o_o$           &       fixed      &   0     &    0    &    all  \\ \hline 
$\rho$(\tBz,\dmd)& &0.007 &$-$0.127 &  \\ \hline \hline

\end{tabular} }
\ec 
\end{table*}

\begin{figure}[t]
\begin{center}
\hs{-0.4cm}\includegraphics[width=9cm,height=11.25cm]{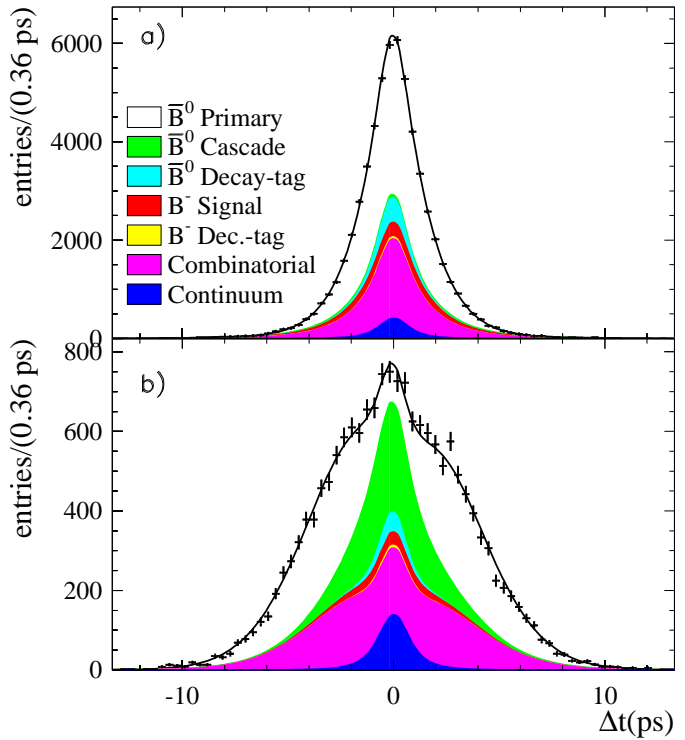}
\end{center}
\caption{Distribution of \deltat\ for unmixed (a) and mixed (b) events in the signal $\mnusq$
region. The points show the data, the curve is 
the projection of the fit result,
and the shaded areas from bottom to top are the contributions from continuum background, \BB\ combinatorial background, peaking \Bub\ background with decay-side tag, 
peaking \Bub\ background with primary tag, 
signal with decay-side tag, signal with cascade tag, and signal with primary tag. } 
\label{f:fitmass1}
\end{figure}

\begin{figure}[!htb]
\begin{center}
\hs{-0.4cm}\includegraphics[width=9cm,height=11.25cm]{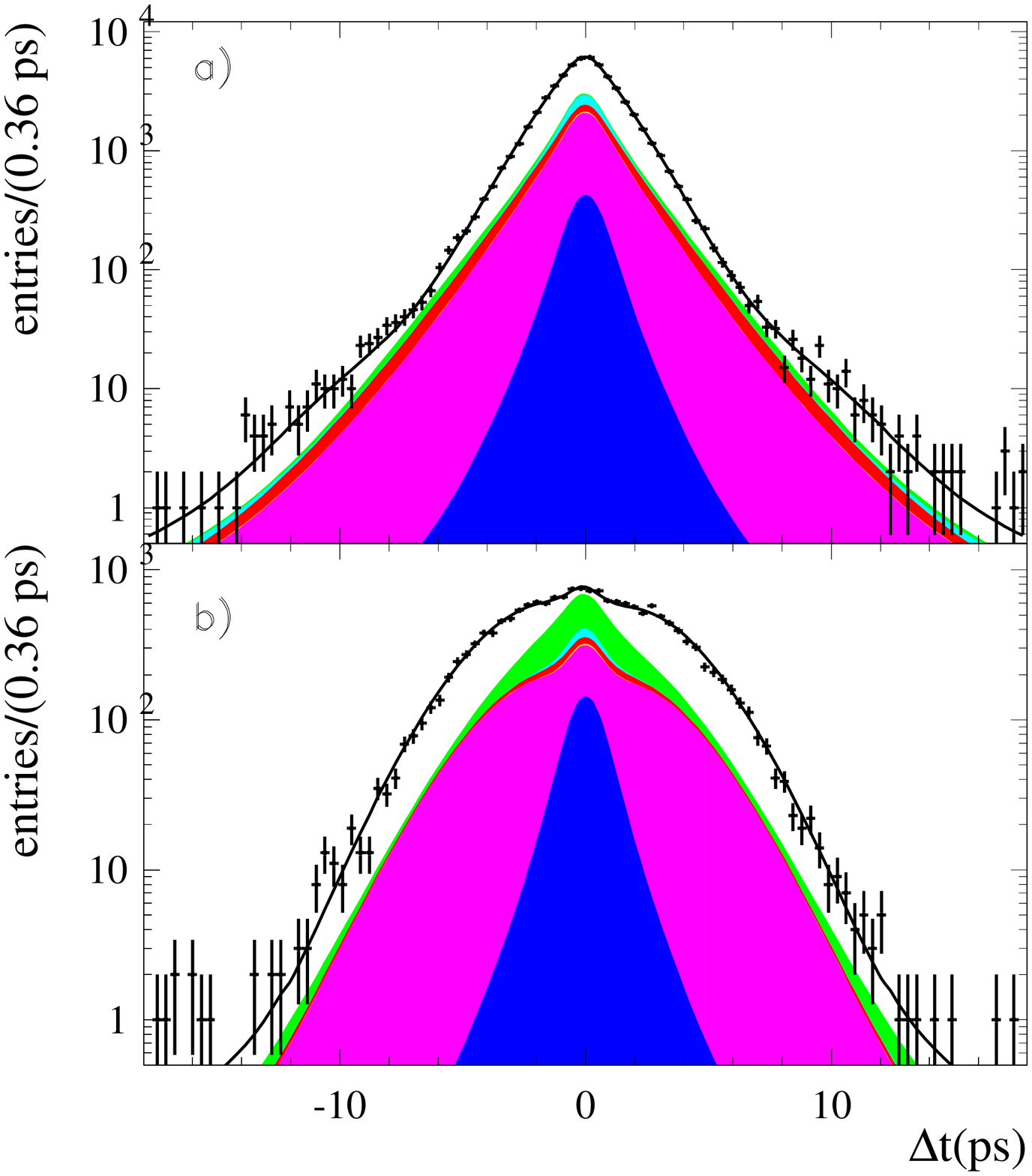}
\end{center}
\caption{Same as Fig.~\ref{f:fitmass1} with logarithmic scale.} 
\label{f:fitmass2}
\end{figure}

\begin{figure}[!htb]
\begin{center}
\hs{-0.4cm}\includegraphics[width=9cm,height=11.25cm]{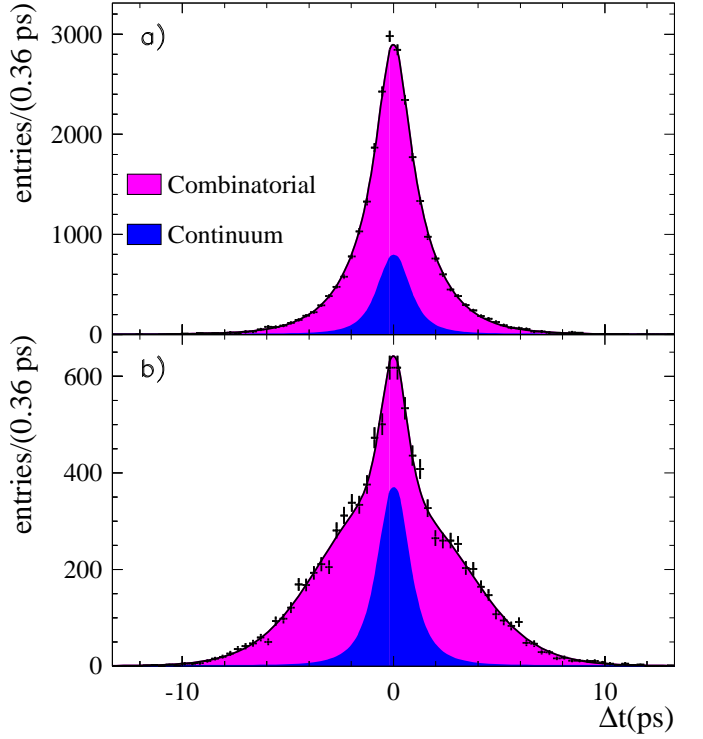} 
\end{center} 
\caption{Distribution of \deltat\ for unmixed (a) and mixed (b) events in the background 
$\mnusq$
region. The points show the data, the curve is 
the projection of the fit result, and
the shaded areas are the contributions from continuum and \BB\ combinatorial background.  } 
\label{f:fitside1}
\end{figure}

\begin{figure}[!htb]
\begin{center}
\hs{-0.4cm}\includegraphics[width=9cm,height=11.25cm]{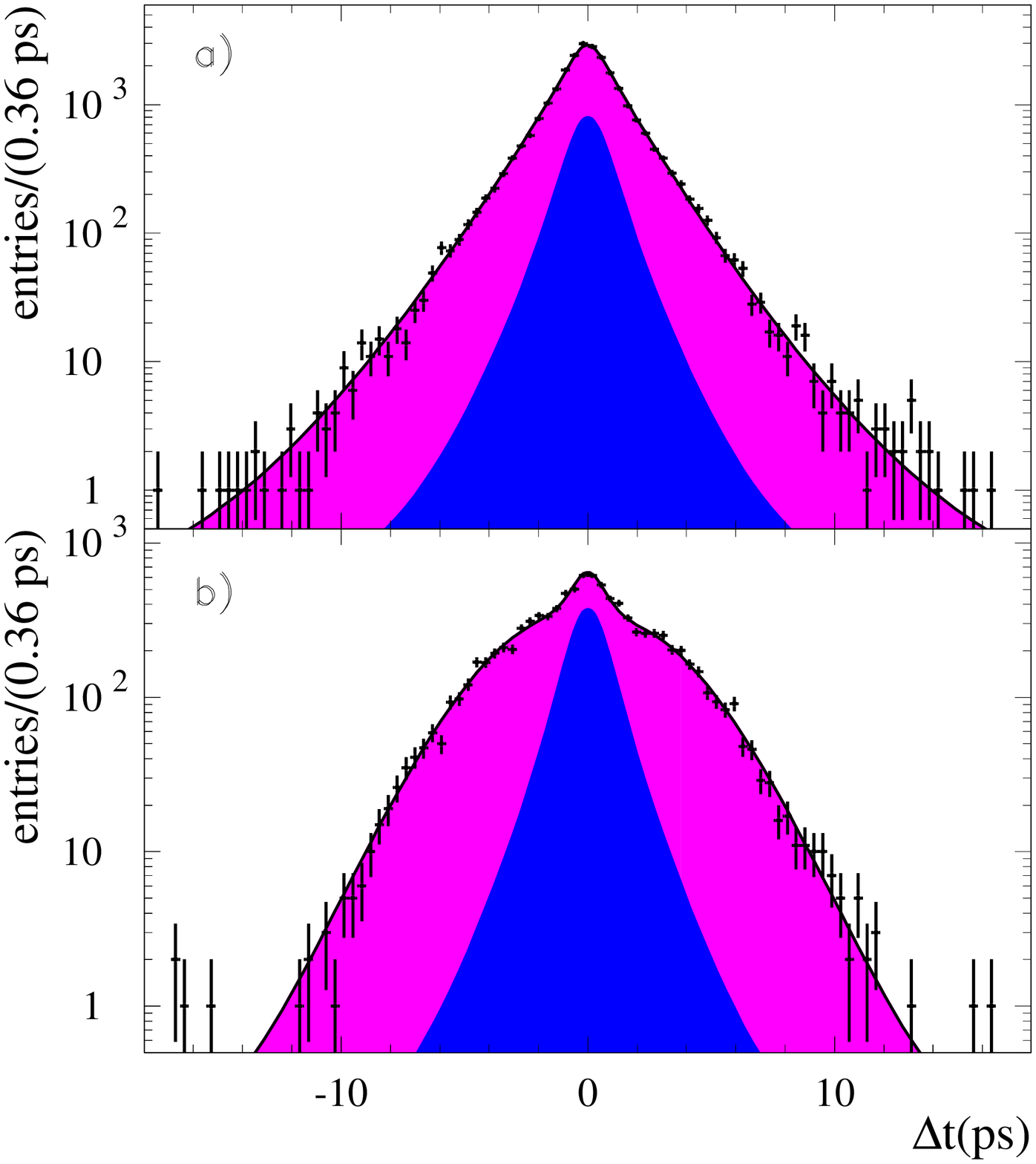}
\end{center} 
\caption{Same as Fig.~\ref{f:fitside1} with logarithmic scale.} 
\label{f:fitside2}
\end{figure}

Details on the systematic error are reported in Sec.~\ref{sec:Systematics}. Figures~\ref{f:fitmass1}
and \ref{f:fitmass2}
show the comparison between the data and the fit function projected on \deltat ,
for a sample of events enriched in signal by the cut  $\mnusq > -2.5$~GeV$^2/c^4$;
Figs.~\ref{f:fitside1} and \ref{f:fitside2} show the same comparison for events in the background region.


\begin{figure}[!htb]
\begin{center}
\hs{-1.5cm}\includegraphics[width=9cm,height=9cm]{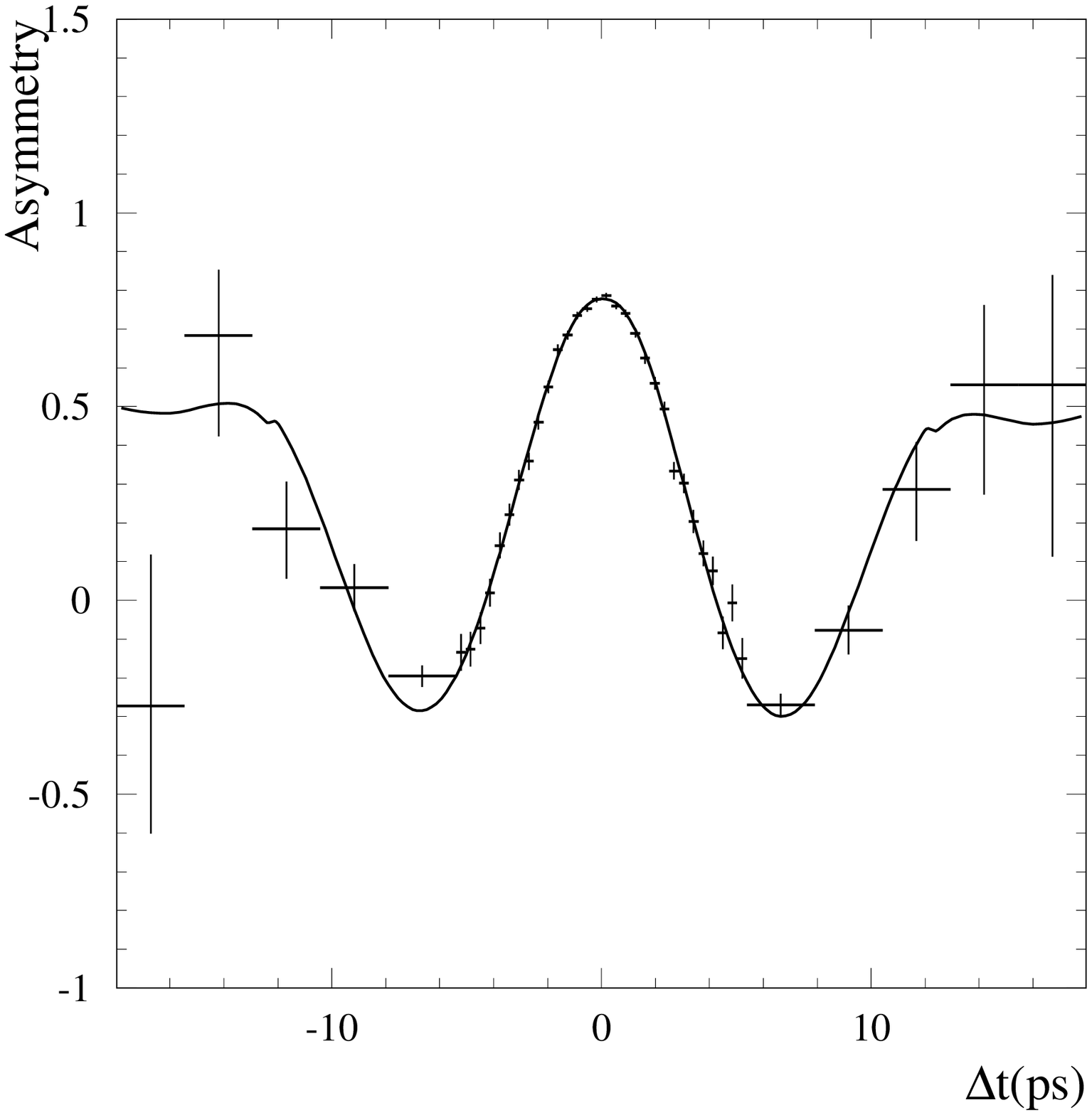} 
\end{center}
\vs{-1.cm}
\caption{Asymmetry between unmixed and mixed events as a function of \deltat,
for events in the signal $\mnusq$ region. Points with error bars represent the
data, and the curve is a projection of the fit result.}
\label{f:asy1}
\end{figure}

\begin{figure}[t]
\begin{center}
\hs{-1.5cm}\includegraphics[width=9cm,height=9cm]{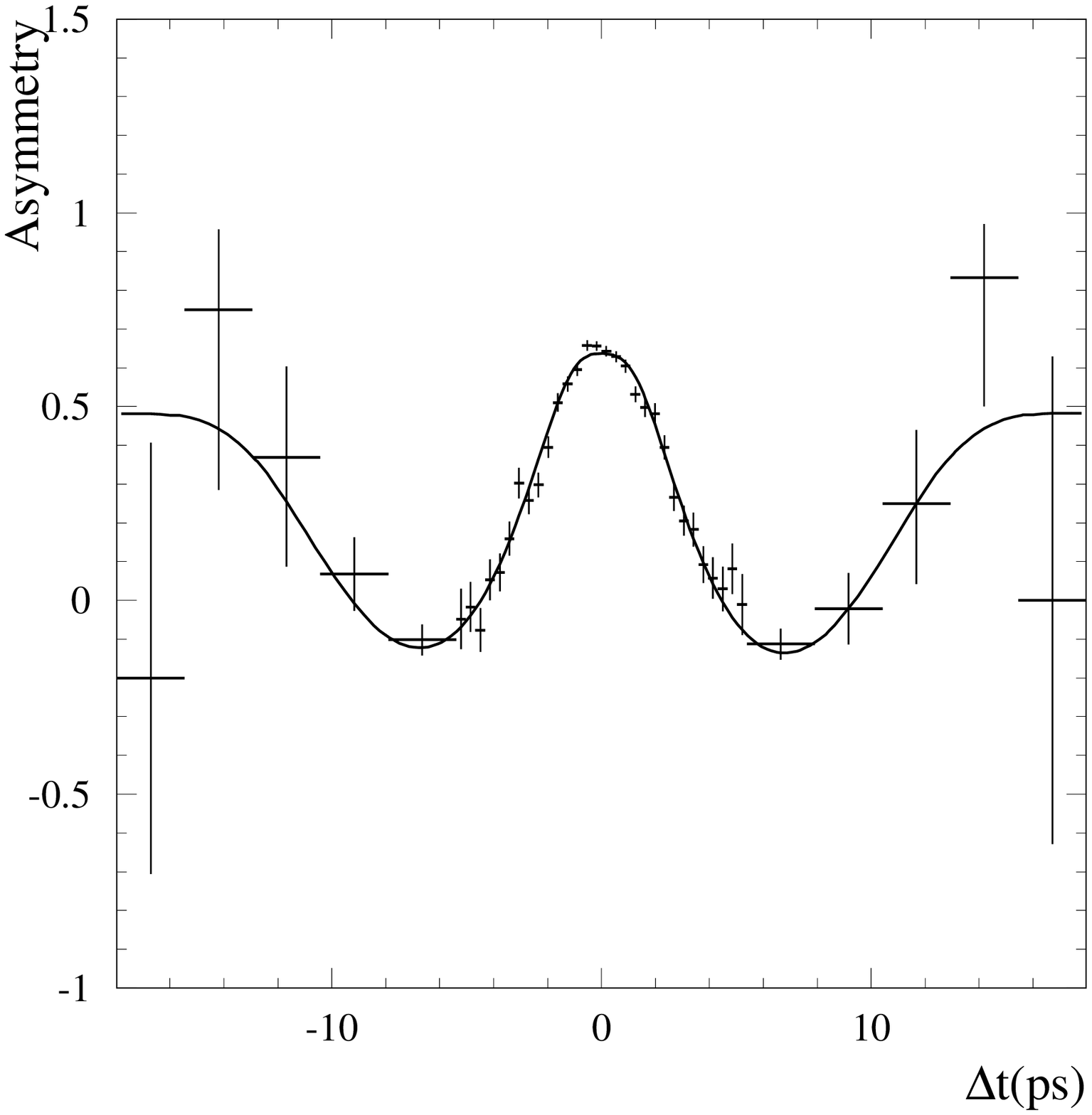}
\end{center}
\vs{-1.cm}
\caption{Asymmetry between unmixed and mixed events as a function of \deltat,
for events in the background $\mnusq$ region. Points with error bars represent the
data, and the curve is a projection of the fit result.}
\label{f:asy2}
\end{figure}

Figures~\ref{f:asy1} and \ref{f:asy2} show plots of the time-dependent asymmetry
\ba
\nonumber 
{\cal A}(\deltat)=\frac{{\cal N}_{\mathrm{unmix}}(\deltat)-{\cal N}_{\mathrm{mix}}(\deltat)}{{\cal N}_{\mathrm{unmix}}(\deltat)+{\cal N}_{\mathrm{mix}}(\deltat)}
\ea
for events in the $\mnusq$ signal region and events in the $\mnusq$ background region.
For signal events, neglecting \deltat\ resolution, ${\cal A}(\deltat) = {\cal D} \cos(\deltamd \deltat)$ (see Eq.~\ref{eq:pdf}).
\par
The agreement between the fit function and the data distribution is good in both the signal and background regions. 
The asymmetry is quite significant for events in the background $\mnusq$ region because a large fraction of these events 
are due to combinatorial \BzBzb\ background.

\section{SYSTEMATIC UNCERTAINTIES}
\label{sec:Systematics}
The systematic errors are summarized in Table~\ref{t:syst}.
We consider the following sources of systematic uncertainty:
\ben
\item Sample composition: 
We calculate a total uncertainty of $\pm 1.3\%$ on the number of signal
events. This uncertainty is the quadratic sum
of the statistical error in the \mnusq\ fit ($\pm$1.2\%), the systematic uncertainty 
on the shape of $\BB$ combinatoric background from the test on the ``wrong-charge'' sample
($\pm 0.2\%$) (see sec. \ref{s:sele}), and the additional systematic uncertainty due to 
low-momentum pions from $D^+$ decays ($\pm 0.4\%$) (see sec. \ref{s:sample}).
\item Analysis bias (entry b): We use the statistical error on the bias observed in the fit on the 
Monte Carlo sample.
\item Signal and background PDF description: Most of the parameters in the PDF are free in the fit
and therefore do not contribute to the systematic error. We vary the parameters that are fixed in the 
fit by their uncertainty, repeat the fit, and use the corresponding variation in $\tBz$ and \dmd\ 
as systematic errors.
We take the uncertainty on \tBu (entry c), and on $D_{\ctl}$ (entry d) from the PDG~\cite{ref:PDG}. 
We find that four parameters used in the description of the combinatorial background, as 
determined by the fit on the Monte Carlo sample, are not in agreement with the Monte Carlo truth. 
They are the fraction of cascade tag-side leptons in the 
unmixed event sample, $f_{\ctl, u}^{BKG}$, the fraction of decay-side tags in the mixed \Bzb and the \Bub 
event samples, $\alpha_{B0, m}^{BKG}$ and $\alpha_{B-, m}^{BKG}$, respectively, 
and an additional parameter used in the description of the shape of the proper time 
difference \deltat\ of the decay-side tagged mixed sample, $f_{\cdl,2}$. 
Therefore we fix them to the Monte Carlo prediction. 
We vary the value of each of them by $20\%$ to compute the systematic error from the comparison
with the default result, and we sum 
the four uncertainties in quadrature (entry e). 
\item Detector alignment: We consider effects due to the detector $z$ scale, determined by 
reconstructing protons scattered from the beam pipe and comparing the measured beam 
pipe dimensions with the optical survey data~\cite{ref:zsc}. The $z$ scale indetermination 
corresponds to an uncertainty of $\pm0.4\%$ on $\deltat$. We repeat the
fit applying this scale correction to $\deltat$, and use the variation with
respect to the default result as the systematic error (entry f). 
From the measurement of the beam energies, the $\Upsilon(4S)$ Lorentz boost factor is determined
with an uncertainty which translates into a $\pm0.1\%$ indetermination on $\deltat$.
Again we repeat the fit and assume as systematic error the variation of the result (entry g). 
We then consider the effect of varying the beam-spot position
by $\pm40~\mu$m in the $y$ direction (entry h). 
We compute the uncertainty due to SVT time-dependent misalignment by comparing results obtained with
different sets of alignment constants (entry i).
\item Decay-side tags: 
We vary the parameters describing the fraction of decay-side tags by their statistical errors,
repeat the fit, and take the variation with respect to the default result as
the systematic error (entry j).
\item Binned fitting: We vary the number of bins in \deltat\ from 100 to 250 
and in \st\ from 20 to 50, and we repeat the fit. 
We take the systematic error to be the variation with respect to the default result (entry k).
\item Outlier description: We vary the value of the offset of the outlier Gaussian  
from $-5~{\rm ps}$  to $+5~{\rm ps}$. As a cross-check, we use a PDF that is uniform in \deltat\ 
for the description of the outliers.
We take the maximum variation with respect to the default result as the systematic error (entry l).
\item Fit range: We vary the \deltat\ fit range from $\pm$18~{\rm ps} to $\pm$ 10~{\rm ps} and the \st\ 
maximum value from 1.8~{\rm ps}
to 4.2~{\rm ps}. Again we assume the maximum variation between the various results and
the default one as the systematic error (entry m).
\item Cascade lepton tag-side parameterization: For the resolution model, we use a Gaussian distribution 
convolved with a one-sided exponential to describe the core part of the resolution function (GExp) 
instead of the Gaussian resolution with a non-zero offset. We quote as systematic error
the difference between the results obtained with the two different approaches (entry n).
\bt[h] \bc 
\caption{\label{t:syst} Systematic uncertainties. See text for details.}
\begin{tabular}{|l|c|c|c|c|}
\hline
Source  & Variation & $\delta\tBz$~(ps) & $\delta\dmd$~(ps$^{-1}$) \\ \hline \hline
\vs{-.3cm}
                             &                &                &               \\ 
(a) Sample Composition       &  $\pm 1.3\%$   &   $\pm0.0003$  & $\mp0.0002$   \\
(b) Analysis bias            &       -        &   $\pm0.0070$  & $\mp0.0035$   \\
(c) \tBu\                    & 1.671$\pm$0.018&   $\mp0.0014$  & $\mp0.0008$   \\  
(d) ${\cal D_{\ctl}}$        & 0.65$\pm$0.08  &   $\mp0.0003$  & $\mp0.0003$  \\ 
(e) Combinatorial BKG        &     -          &   $\pm 0.0007$  &  $\mp0.0002$     \\ 
(f) $z$ scale                &     -          &   $\pm 0.0070$  &  $\mp0.0020$     \\ 
(g) PEP-II boost             &     -          &   $\pm 0.0020$  &  $\mp0.0003$     \\ 
(h) Beam-spot position       &     -          &   $\pm 0.0050$  &  $\mp0.0010$     \\  
(i) Alignment                &     -          &   $^{+0.0132}_{-0.0038}$  &  $^{-0.0038}_{+0.0033}$     \\ 
(j) Decay-side tags          &     -          &   $\pm 0.0013 $ &  $ - $     \\ 
(k) Binning                  &     -          &   $\mp 0.0021 $ &  $\pm0.0006$     \\ 
(l) Outlier parameters       &     -          &   $\pm 0.0028 $ &  $\pm0.0012$     \\ 
(m) \deltat\ and \st\ cut    &     -          &   $\pm 0.0033 $ &  $\mp0.0033$     \\ 
(n) GExp model               &     -          &   $-0.0016    $ &  $+0.0011$     \\ 
\hline
Total                        &                   &   $^{+0.0182}_{-0.0131}$  &  $^{+0.0068}_{-0.0064}$    \\ \hline \hline
\end{tabular}
\ec \et
\een

\section{CONSISTENCY CHECKS}
We rely on the assumption that the parameters of the background PDF do not depend on $\mnusq$.
We verify this assumption for the continuum background with the fit to the off-resonance events. 
To check this assumption for the \BB\ combinatorial PDF, we perform several cross checks on the data and the Monte Carlo.
We compare the simulated combinatorial \BB \deltat\ distribution in several independent 
regions of $\mnusq$ 
with Kolmogorov-Smirnov tests and always obtain a reasonable probability for agreement. We fit 
the \deltat\ distribution of combinatorial background \BB\ events 
separately in the signal and background $\mnusq$ region and compare the parameters of the PDF.
We fit the signal plus \BB\ background Monte Carlo events in the signal region only,
fixing all the parameters of the \BB\ background to the values obtained in a fit to the 
background region, and do not
see any significant deviation from the results of the full fit.
Finally, we repeat 
the fit on both the data and the Monte Carlo using different $\mnusq$ ranges for the background region.
Once  again, we do not observe any significant difference in $\tBz$ and \dmd\ relative to the default 
result.
\par
We repeat the analysis with a more stringent requirement on the combined signal likelihood
(a minimum ${\cal X}$ of 0.5 rather than 0.4). No significant change in the result is observed.
\par
We validate the fit procedure with a parameterized Monte Carlo simulation. 
We simulate several experiments from the
fitted PDF of both the Monte Carlo and the data, with parameters fixed to the
values obtained from the corresponding fit.
Each experiment is produced with the same number of events as the original sample. 
For each experiment we produce seven data sets, corresponding to $\Bzb$ with primary, cascade, 
and decay-side lepton tags, peaking $\Bub$ background with tag-side and decay-side lepton tags,
$\BB$ combinatorial background, and continuum background. We fit every experiment with the
same procedure as the corresponding original sample, and finally we compare the fitted
parameters with the generated values. 
The result of this study is summarized in Table~\ref{t:toy} where we report the average
and the root-mean-square deviation (rms) of the distribution of the difference between the
fitted and the generated parameter value divided by the fit statistical error (pull). 
We do not find any significant statistical anomaly in the fit behaviour.
\bt[h] \bc 
\caption{\label{t:toy} Results extracted from parameterized Monte Carlo experiments generated 
with parameters fixed to the values obtained from the fit to data (second column) 
and to the full Monte Carlo simulation (third column). For both $\tBz$ and $\dmd$
the average and the rms of the distribution of the pull with respect to the generated value
are reported. 
}
\begin{tabular}{|l|c|c|}
\hline
Parameter & data   & Monte Carlo \\ \hline \hline
\vs{-.3cm}
                             &                &               \\ 
number of experiments   & 54            & 124 \\
Pull $\tBz$ average	& $0.38\pm0.19$ & $0.38\pm 0.12$ \\
Pull $\tBz$ rms 	& $1.13\pm0.19$ & $1.25\pm 0.13$ \\
Pull $\dmd$ average	& $-0.33\pm0.17$ & $0.08\pm 0.11$ \\
Pull $\dmd$ rms 	& $1.14\pm0.17$ & $1.09\pm 0.08$ \\
\hline \hline
\end{tabular}
\ec \et

We rely on the assumption that the decay-rate difference $\Delta \Gamma_d$ between the two 
mass eigenstates can be neglected in the analysis. 
We check this assumption with a parameterized
Monte Carlo simulation in which events are simulated with zero mistag probability and perfect \deltat\ resolution.
We produce two sets of one hundred Monte Carlo experiments. In the first set, 
$\Delta \Gamma_d=0$; in the second, $\frac{\Delta \Gamma_d}{\Gamma_d}=0.01$.
We fit every experiment with the same procedure neglecting $\Delta \Gamma_d$ 
and we do not find any significant difference in the values of $\tBz$ and \dmd\ in
the two different sets.

We investigate a possible analysis bias due to the finite $\tau$ and $D_s$
lifetimes in
$\Bzb \ra \dsp \tau^- \bar{\nu}_\tau $ ($\tau^- \ra \ell^- X$) and 
$\Bzb \ra \dsp D_s^- $ ($D_s^- \ra \ell^- X$) decays.
We fit the Monte Carlo signal sample 
with no mistag and realistic \deltat\ resolution
after removing these decays and we do not find any 
significant variation with respect to the result obtained with the full signal sample.

\section{CONCLUSION}
\label{sec:Summary}
We have performed a measurement of \dmd\ and $\tBz$ with a sample of about
50\,000 partially reconstructed, lepton-tagged \BtoDs\ decays. We obtain the
following results:
\ba
\nonumber \tBz &=& (1.504 \pm 0.013 ~\mathrm{(stat)}~^{+0.018}_{-0.013} ~\mathrm{(syst))~ps}, \\
\nonumber \dmd &=& (0.511 \pm 0.007 ~\mathrm{(stat)}~^{+0.007}_{-0.006} ~\mathrm{(syst))~ps}^{-1}.
\ea
The $\tBz$ value is consistent with the published measurement performed by 
\babar\ using \BtoDs\ partially reconstructed decays~\cite{ref:t1}.
Our results are also consistent with published measurements 
of $\tBz$ and $\dmd$ 
performed by \babar\ with different data sets~\cite{ref:t2,ref:dilepton,ref:xl,ref:hdmd,ref:htau}, 
and with the world averages computed
by the Heavy Flavor Averaging Group for the PDG 2005 web update: 
$\tBz = (1.532 \pm 0.009)$~ps, and
$\dmd = (0.505 \pm 0.005)$~ps$^{-1}$.

\section{ACKNOWLEDGMENTS}
\label{sec:Acknowledgments}
\input pubboard/acknowledgements

\end{document}

%% file: pubboard/authors_may2005.tex
%
\author{B.~Aubert}
\author{R.~Barate}
\author{D.~Boutigny}
\author{F.~Couderc}
\author{Y.~Karyotakis}
\author{J.~P.~Lees}
\author{V.~Poireau}
\author{V.~Tisserand}
\author{A.~Zghiche}
\affiliation{Laboratoire de Physique des Particules, F-74941 Annecy-le-Vieux, France }
\author{E.~Grauges}
\affiliation{IFAE, Universitat Autonoma de Barcelona, E-08193 Bellaterra, Barcelona, Spain }
% ...
\collaboration{The \babar\ Collaboration}
\noaffiliation

%% file: pubboard/acknowledgements.tex
We are grateful for the 
extraordinary contributions of our \pep2\ colleagues in
achieving the excellent luminosity and machine conditions
that have made this work possible.
The success of this project also relies critically on the 
expertise and dedication of the computing organizations that 
support \babar.
The collaborating institutions wish to thank 
SLAC for its support and the kind hospitality extended to them. 
This work is supported by the
US Department of Energy
and National Science Foundation, the
Natural Sciences and Engineering Research Council (Canada),
Institute of High Energy Physics (China), the
Commissariat \`a l'Energie Atomique and
Institut National de Physique Nucl\'eaire et de Physique des Particules
(France), the
Bundesministerium f\"ur Bildung und Forschung and
Deutsche Forschungsgemeinschaft
(Germany), the
Istituto Nazionale di Fisica Nucleare (Italy),
the Foundation for Fundamental Research on Matter (The Netherlands),
the Research Council of Norway, the
Ministry of Science and Technology of the Russian Federation, and the
Particle Physics and Astronomy Research Council (United Kingdom). 
Individuals have received support from 
CONACyT (Mexico),
the A. P. Sloan Foundation, 
the Research Corporation,
and the Alexander von Humboldt Foundation.

%% file: paper_eprint.bbl
\begin{thebibliography}{99}
\bibitem{ref:footnote1}
Charge conjugate states are always implicitly assumed; $\ell$ means either electron
or muon.
\bibitem{ref:footnote2}
Throughout the paper the momentum, energy and 
direction of all particles    
are computed in the \epem\ rest frame.
\end{thebibliography}
